\documentclass[twocolumn]{aastex62}
\usepackage{graphicx}
\usepackage{xcolor}
\usepackage{threeparttable}
\hypersetup{linkcolor=purple,citecolor=xlinkcolor,filecolor=xlinkcolor,urlcolor=xlinkcolor}
\shorttitle{Most Massive Galaxies at $3<z<6$ selected from UltraVISTA Deep-Stripes}
\shortauthors{Marsan et al.}

\begin{document}


\title{The Number Densities and Stellar Populations of Massive Galaxies at $3 < z < 6$: 
A Diverse, Rapidly Forming Population in the Early Universe}

\correspondingauthor{Z. Cemile Marsan}
\email{cmarsan@yorku.ca}

\author[0000-0002-7248-1566]{Z. Cemile Marsan}
\altaffiliation{York Science Fellow}
\affil{Department of Physics and Astronomy, York University, 4700, Keele Street, Toronto, ON MJ3 1P3, Canada}

\author[0000-0002-9330-9108]{Adam Muzzin}
\affil{Department of Physics and Astronomy, York University, 4700, Keele Street, Toronto, ON MJ3 1P3, Canada}

\author[0000-0001-9002-3502]{Danilo Marchesini}
\affil{Department of Physics and Astronomy, Tufts University, 574 Boston Avenue Suites 304, Medford, MA 02155, USA}

\author[0000-0001-7768-5309]{Mauro Stefanon}
\affil{Leiden Observatory, Leiden University, 2300 RA Leiden, The Netherlands}

\author[0000-0003-3243-9969]{Nicholas Martis}
\affil{Department of Physics and Astronomy, Tufts University, 574 Boston Avenue Suites 304, Medford, MA 02155, USA}

\author[0000-0002-8053-8040]{Marianna Annunziatella}
\affil{Department of Physics and Astronomy, Tufts University, 574 Boston Avenue Suites 304, Medford, MA 02155, USA}
\affil{Centro de Astrobiolog\'ia (CSIC-INTA), Ctra de Torrej\'on a Ajalvir, km 4, E-28850 Torrej\'on de Ardoz, Madrid, Spain}

\author[0000-0001-6251-3125]{Jeffrey C.~C. Chan}
\affil{Department of Physics and Astronomy, University of California, Riverside, 900 University Avenue, Riverside, CA 92521, USA}

\author[0000-0003-1371-6019]{Michael C. Cooper}
\affil{Center for Cosmology, Department of Physics and Astronomy, University of California, Irvine, 4129 Frederick Reines Hall, Irvine, CA,USA}

\author[0000-0001-6003-0541]{Ben Forrest}
\affil{Department of Physics and Astronomy, University of California, Riverside, 900 University Avenue, Riverside, CA 92521, USA}

\author[0000-0003-0408-9850]{Percy Gomez}
\affil{W.M. Keck Observatory, 65-1120 Mamalahoa Hwy., Kamuela, HI 96743, USA}

\author{Ian McConachie}
\affil{Department of Physics and Astronomy, University of California, Riverside, 900 University Avenue, Riverside, CA 92521, USA}

\author[0000-0002-6572-7089]{Gillian Wilson}
\affil{Department of Physics and Astronomy, University of California, Riverside, 900 University Avenue, Riverside, CA 92521, USA}



\begin{abstract}

We present the census of massive (log(M$_{*}$/M$_{\odot}$)~$> 11$)  galaxies at $3<z<6$
identified over the COSMOS/UltraVISTA Ultra-Deep field stripes:~
consisting of $\approx100$ and $\approx20$ high-confidence candidates at $3<z<4$ and $4<z<6$, respectively.
The $3<z<4$ population is comprised of post-starburst, UV star-forming and dusty-star forming galaxies in roughly equal fractions, 
while UV-star-forming galaxies dominate at $4<z<6$ . 
We account for various sources of biases in SED modelling, finding that the treatment of emission line contamination
is essential for understanding the number densities and mass growth histories of massive galaxies at $z>3$. 
The significant increase in observed number densities at $z\sim4$ ($>\times$~5 in $\lesssim600$~Myrs) implies that 
this is the epoch at which log(M$_{*}$/M$_{\odot}$)~$> 11$ galaxies emerge in significant numbers, 
with stellar ages ($\approx500-900$~Myrs) indicating rapid formation epochs as early as $z\sim7$.
Leveraging ancillary multi-wavelength datasets, we perform panchromatic SED modelling to constrain the total star-formation activity of the sample. 
The star-formation activity of the sample is generally consistent with being on the 
star-formation main sequence at the considered redshifts, with $\approx15-25\%$ of the population showing evidence of suppressed star-formation rates, 
indicating that quenching mechanisms are already at play by $z\sim4$. 
We stack available \textit{HST} imaging, confirming their compact nature 
($r_{e}\lesssim2.2$~kpc), consistent with expected sizes of high-$z$ star-forming galaxies. 
Finally, we discuss how our results are in-line with the early formation epochs and short formation timescales inferred from the fossil records of the most massive galaxies in the Universe. 

\end{abstract}


\section{Introduction} \label{sec:intro}
In the nearby universe, the most massive galaxies are a relatively homogenous population: with early-type morphologies, little dust attenuation, and quiescent, metal-rich old ($>10$~Gyr) stellar populations \citep{gallazzi05, gallazzi06, thomas05, mcdermid15, choi14, bellstedt20}. 
This implies that the majority of their stars were formed in the first few Gyr of cosmic history 
(i.e., $z> 2$) through short and intense, bursts of star formation (e.g., \citealt{renzini06,vandokkum07,thomas10}). 
Indeed, numerous works have demonstrated that old, passively evolving galaxies exist at $1.5<z<2.5$ \citep{cimatti04, cimatti08, daddi05, kriek09a, whitaker13,belli17b, belli19, toft17, newman18, carnall19a}, while recent ambitious spectroscopic campaigns have confirmed the existence of massive, evolved galaxies at 
$3<z<4$ \citep{marsan15, marsan17, glazebrook17, schreiber18b, tanaka19, forrest20a, forrest20b, valentino20, saracco20}.
Related to this, the number density of the most massive galaxies show no evolution in the $\approx3.5$~Gyrs between $z=0.4$ and $z\sim1-1.5$ \citep{kawinwanichakij20}, and evolves very little in the prior $\sim3-4$ Gyr to $z\approx4$, providing additional evidence that they must have assembled important amounts of their stellar content rapidly beyond $z>3$ \citep{franx03, cimatti04, daddi04, perezgonzalez08, wiklind08, marchesini10, brammer11, caputi12, ilbert13, muzzin13b, duncan14,
nayyeri14, straatman14,  tomczak14,  grazian15, davidzon17}. 

Massive galaxies display more diverse stellar populations with increasing cosmic look-back time: at $z\sim2$, half of massive galaxies are already devoid of star formation and have old stellar ages suggesting that they quenched their star formation at even earlier times (e.g. \citealt{kriek06, franx08, toft09, mccracken10, wuyts11, brammer11, whitaker11, kadofong17, morishita18, deugenio20}), while at $z\sim3$ the population is dominated by dusty, star-forming galaxies \citep{marchesini10, spitler14, martis16, martis19, deshmukh18}. Massive galaxies in the early Universe are also more compact than their local counterparts \citep{daddi05, vanderwel14a, straatman15, mowla19}, with a more dramatic size evolution observed for the quiescent population, \textit{e.g}., \citealt{vandokkum08b, williams10, newman12}. The resolved light distributions for the most massive galaxies at $z\sim1.5-3$ show range in morphologies with significant fractions ($\sim30-40\%$) exhibiting features indicative of interactions and /or companions in close proximity \citep{marsan19, stockmann20}. 

Deeper extragalactic surveys constructed in recent years have uncovered  
a substantial population of massive, evolved galaxies exists beyond $z>3$, although statistical samples are still missing. 
The discovery of massive, evolved galaxy candidates at $z\gtrsim4$ have been claimed by several works \citep{stefanon15, caputi15, mawatari16, mawatari19, wang16, wang19, merlin18, merlin19, alcalde19, girelli19, santini19, guarnieri20, carnall20}, however, these largely rely on detections in relatively few photometric data points. 
Thus, the task of determining when the first massive galaxies emerge, and understanding the baryonic physics responsible for their accelerated evolution in the first billion years of the universe is still incomplete. 

In this study, we aim to 
study the most massive galaxies present in the first $\sim2.5$Gyrs of the Universe using a statistically large sample, and extend the investigation of the most massive galaxies even further at $z>4$, the epoch at which these extreme systems are expected to emerge. 
While this has been examined over smaller fields by previous works 
(\textit{e.g.}, \citealt{mobasher05}, \citealt{wiklind08}, \citealt{fontana09}, \citealt{nayyeri14}), this work takes advantage 
of the unique combination of deep near- and mid- infrared observations over a wide area ($\approx0.84$~deg$^2$ in the COSMOS field) 
provided by the UltraVISTA and SMUVS surveys, crucial to selecting statistical samples of evolved galaxies in the early universe (also see \citealt{caputi15}).
The COSMOS field benefits from a multitude of (deep) ancillary datasets collected across the electromagnetic spectrum. 
Particularly relevant for understanding the massive high-$z$ galaxy population, 
we utilize the available far-infrared, radio and X-ray datasets, as well as imaging data from the Hubble Space Telescope (\textit{HST}), to identify multi-wavelength counterparts and explore
the prevalence of AGN, the level of star-formation activity obscured by dust, as well as size estimates.

This paper is structured as follows:
Section~\ref{sec:data} describes the parent DR3 galaxy catalog and the quality cuts employed to identify the \textit{bona fide} mass complete (log(M$_{*}$/M$_{\odot}$)~$> 11$) sample of galaxies at $3<z<6$, as well as the ancillary multi-wavelength datasets incorporated in our analyses. 
We dissect and characterize the stellar populations and AGN fraction of the sample in Section~\ref{sec-charpresent}. 
Tests for the robustness of stellar-mass estimates and the high-confidence number density measurements of massive galaxies 
at $3<z<6$ are presented in Section~\ref{sec-abundances}. 
Section~\ref{sec-sfactivity} explores the star-formation activity of the sample via the modelling of their extended UV-FIR/radio SEDs. 
The sizes of high-z massive galaxies are investigated by the stacking analysis of available \textit{HST} imaging in Section~\ref{sec-size}. 
In Section~\ref{sec:disc} we discuss the implications of our results in the context of formation mechanisms and evolutionary pathways inferred for the most massive galaxies in the universe. Summary is presented in Section~\ref{sec:summary}.
The standard $\Lambda$CDM cosmological parameters $\Omega_{M} = 0.3$, $\Omega_{\Lambda} = 0.7$ 
with $H_{0} = 70~{\rm{km~s}}^{-1} {\rm{Mpc}}^{-1}$ and a \citet{chabrier03} initial mass function (IMF) are assumed 
throughout this paper. All magnitudes are in the AB system. 

\section{Data and Sample}\label{sec:data}
This analysis is based on the $K_{\rm S}$-selected catalog constructed over the UltraVISTA near-infrared imaging survey \citep{mccracken12} in the COSMOS \citep{scoville07} field, covering an area totalling 
$\approx0.84~$deg$^2$ in the four strips of \textit{ultra-deep} UltraVISTA imaging (Data Release 3), where both IRAC and optical coverage is available (covering the three main \textit{deep} strips and half of the final \textit{deep} strip).
Briefly, the DR3 photometric catalog used in this work was constructed using the same procedure as in \citet{muzzin13a}, but the NIR depths are $\approx1.2$ mag deeper compared to the DR1 release, reaching $Y=25.8$, $J=25.6$, $H=25.4$ and $K_{\rm {S}}=25.2$ mag ($5\sigma$, $2.''1$ diameter aperture). 

In Section~\ref{sec:DR3cat}, we summarize the galaxy catalog used 
to obtain the mass-complete galaxy sample at $3<z<6$ ($M_{*}\geq10^{11}M_{\odot}$), and describe the quality cuts employed to arrive at the \textit{bona fide} sample of massive galaxies in Section~\ref{sec-sampsel} (105 and 23 galaxies at $z<4$ and $z>4$, respectively). 
We take advantage of the wealth of photometry across a range of wavelengths 
in the COSMOS field in order to 
reliably constrain the properties of massive galaxies at $3<z<6$ (e.g., determining AGN hosts, constraining dust-obscured SFRs). 
The ancillary multi-wavelength source catalogs used in this analysis and counterpart identification methods are outlined in Section~\ref{sec:counterparts}. 

\subsection{UltraVISTA DR3 Photometric Catalog}\label{sec:DR3cat}

\begin{figure*}
\plotone{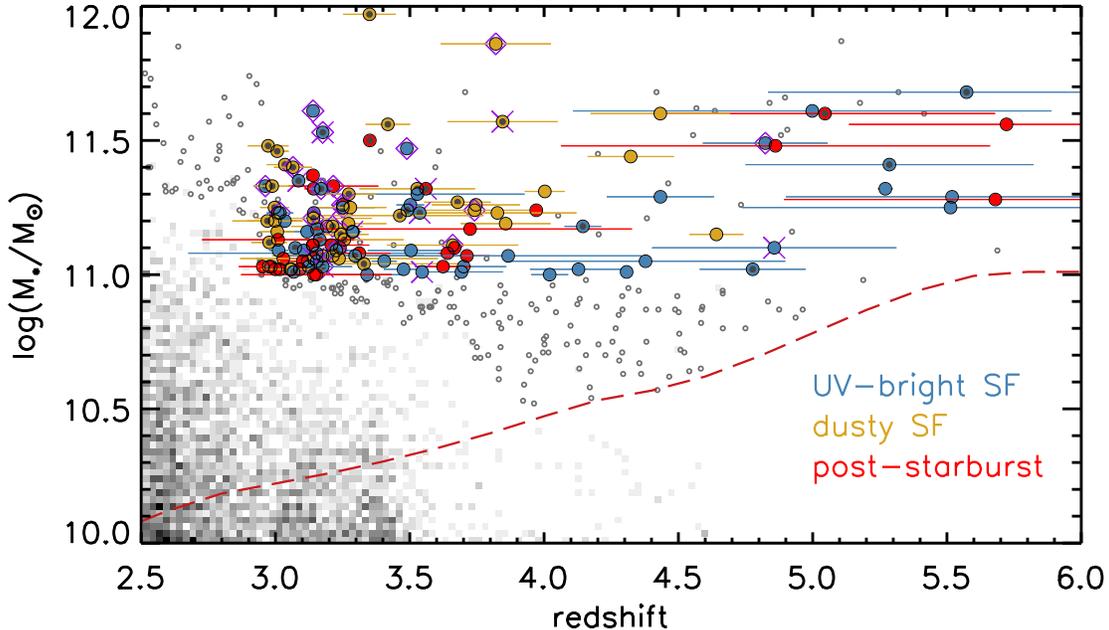}
\caption{Galaxy stellar masses as a function of redshift. 
The grayscale represents galaxies down to the 90$\%$ $K_{\rm {S}}$-band completeness limit of the UltraVISTA DR3 catalog. Empty circles are used to highlight objects in low-density regions 
above the adopted $95\%$ mass-completeness limit, which is shown by the red dashed curve. 
Large symbols indicate the \textit{bona fide} mass-complete (log($M_{*}/M_{\odot}$)$>11.0$) galaxy sample (Section~\ref{sec-sampsel}), 
colored according to the classification scheme in Section~\ref{sec-dissect}: \textit{red}, \textit{blue} and \textit{gold} for post-starburst, 
UV star-forming, dusty star-forming subsamples, respectively. Galaxies in our sample with X-ray and radio identified counterparts are indicated with X and diamond symbols, respectively. 
Small (central) black points indicate those with $>3\sigma$ \textit{Spitzer} 24$\mu$m band detections. 
\label{fig-masscompleteness}}
\end{figure*}

In addition to the 30 photometric bands used in the construction of the UltraVISTA DR1 $K_{\rm S}$-selected catalog presented in \citep{muzzin13a}, we include the ultra-deep optical imaging ($\approx 27.0$~mag AB)
from Subaru Hyper Suprime-Cam (DR1, \citealt{aihara17}). 
Particularly crucial for identifying evolved galaxies at early cosmic epochs, this catalog is complemented by 
full-depth \textit{Spitzer}/IRAC mosaics built combining S-COSMOS \citep{sanders07}, 
the \textit{Spitzer} Large Area Survey with Hyper-Suprime-Cam (SPLASH, \citealt{mehta18}) and 
the \textit{Spitzer} Matching survey of the UltraVISTA ultra-deep Stripes (SMUVS, \citealt{caputi17, ashby18}), 
pushing the available IRAC imaging $\sim1$~mag deeper at  $3.6\mu$m and $4.5\mu$m over the ultra-deep stripes (compared to DR1). The catalog also includes UVISTA NB118, five CFHTLS deep, and NMBS filters \citep{whitaker11}, totalling 49 bands where there is overlap. 
The total $K_{\rm S}$-band $90\%$ completeness limit adopted in this work is $K_{\rm{S}}=24.5$~mag, corresponding to a stellar-mass completeness limit of $M_{*}=10^{11}M_{\odot}$ at $z=6$ (Figure~\ref{fig-masscompleteness}). 

Photometric redshifts ($z_{\rm{phot}}$) are calculated for each entry in the $K_{\rm S}$-selected catalog using the photometric redshift code EAZY \citep{brammer08}. Briefly, EAZY fits the spectral energy distributions (SEDs) using linear combinations of template SEDS. 
The template set employed here is
comprised of those derived from the PEGASE models \citep{fioc99}, a red template from the models of \citet{maraston05}, 
a post-starburst \citet{bc03} model, as well as a 
template to account for galaxies that are both old and dusty. Photometric redshifts were determined with EAZY allowing solutions in the range $0<z<6$ using the \texttt{v1.0} template error function and $K_{\rm{S,tot}}$ magnitude prior. 


Similar to \citet{muzzin13a}, stellar population parameters for catalog sources were estimated using the FAST code \citep{kriek09a} to fit galaxy SEDs with \citet{bc03} stellar population synthesis models. We assume solar metallicity, a \citet{chabrier03} initial mass function (IMF), and the
\citet{calzetti00} dust extinction law. The set of template SEDs were constructed adopting exponentially declining star formation histories (SFHs) of the form $SFR \propto exp(-t/\tau)$, where $\tau$ is the $e$-folding star formation timescale. We adopted a grid for log($\tau$) between 7.0 and 10.0 in steps of 0.1~dex and for dust attenuation ($A_{V}$) between 0 and 5 mag in steps 0.1 mag. 
The time since the onset of star formation ($t$) was allowed to range between 10~Myrs and the maximum age of the universe at the redshift of sources in steps of 0.1~dex. All galaxies were fit assuming the $z_{\rm phot}$ EAZY solutions, or spectroscopic redshifts when available. 
The stellar masses ($M_{*}$) are determined by multiplying the mass-to-light ratio of the best-fit SED by its corresponding normalization factor. 

Figure~\ref{fig-masscompleteness} shows the stellar mass of galaxies down to the $90\%$ $K_{\rm{S,~tot}}$-band completeness of the survey 
($K_{\rm{S}}=24.5$, in grayscale). We estimate the redshift-dependent mass-completeness limit of the survey following the approach outlined in \citet{marchesini09} and also adopted in \citet{tomczak14}. 
Briefly, we first scaled the stellar mass of each galaxy down to the $K_{\rm S}$-band $90\%$ completeness limit. 
Then, in each redshift bin, the upper envelope below which 95$\%$ of the galaxies are positioned is estimated to be the 95$\%$ mass-completeness limit. 
This empirical $95\%$ mass-completeness limit is indicated as the red dashed curve in Figure~\ref{fig-masscompleteness}. 

The resulting stellar mass completeness limit we adopt in this work is $M_{*}=10^{11}M_{\odot}$ over the targeted redshift range $3.0\leq z<6.0$.
Large colored symbols in Figure~\ref{fig-masscompleteness} indicate the \textit{bona fide} sample of high-redshift massive galaxy candidates which have passed our initial quality assessments 
(summarized in Section~\ref{sec-sampsel}). 

\subsection{Sample Selection}\label{sec-sampsel}

From the DR3 galaxy catalog outlined in \ref{sec:DR3cat}, we
selected all targets with stellar masses ($M_{*}\geq10^{11}M_{\odot}$) and $z_{\rm{phot}}$ consistent with $3<z<6$ (within uncertainties), yielding $\approx180$ objects. 
We visually inspected the NIR image stamps, and excluded targets near bright stars with potential contaminated photometry ($\sim30$ targets). 
Targets with poor fits to the observed photometry were removed after checking the photometric redshift distribution estimated by EAZY and best-fit SEDs obtained using FAST (eliminating another $\sim20$ objects, the majority of which were quasars or diffraction spikes). 
This yields a sample of 105 and 23 galaxies selected at $3<z<4$ and $4<z<6$, respectively. 
We find that some objects in this sample have photometric redshift solutions which hits the $z_{\rm{max}}=6$ boundary, therefore we re-rerun EAZY and FAST 
extending $z_{\rm{max}}=7$. 

\subsubsection{Spectroscopic Redshifts}
The high-$z$ massive galaxy sample includes twelve objects with spectroscopic redshift measurements in the literature. 
The majority of redshifts are obtained using deep NIR spectroscopy: `C1-23152' in \citet{marsan15} at $z_{\rm{spec}}=3.351$,  
`C1-15182' in \citet{marsan17} at $z_{\rm{spec}}=3.370$, 
and `ZF-COSMOS-20115' presented in \citet{glazebrook17} at $z_{\rm{spec}}=3.717$. 
Eight additional objects from the sample of massive galaxies at $3<z<4$ were targeted with follow-up Keck-MOSFIRE 
spectroscopy as part of the MAGAZ3NE Survey, yielding seven $z_{\rm{spec}}>3$ redshift confirmations, 
while one was found to be $z_{\rm{spec}}=2.80$ \citep{forrest20b}. 
Two galaxies have high-quality redshifts from the DEIMOS 10k spectroscopy catalog (\citealt{hasinger18}; 
at $z_{\rm{spec}}=3.086$ and $3.176$; 
six additional galaxies have $z_{\rm{spec}}>3$, although with lower quality flags). 
The redshifts for these twelve objects are fixed to the $z_{\rm{spec}}$ when performing SED fits. 

\subsection{Identifying Counterparts in Ancillary Datasets}\label{sec:counterparts}

\begin{figure}[!htbp]
\begin{center}
\includegraphics[width=\linewidth]{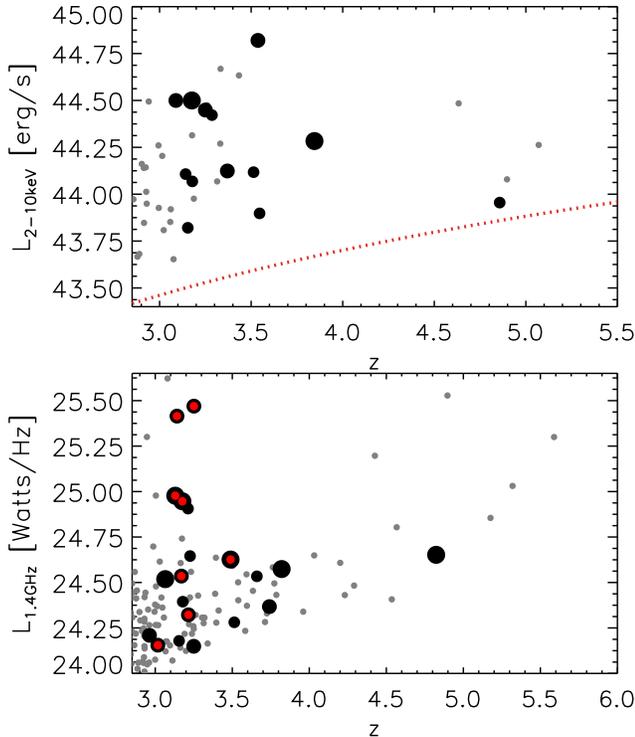}
\caption{Results of cross-matching the $K_{\rm{S}}$-selected UltraVISTA DR3 sample with the Chandra COSMOS Legacy \citep{civano16} and VLA-COSMOS 3GHz Large Project \citep{smolcic17a} source catalogs. 
Targets among the \textit{bona fide} massive galaxy sample with X-ray or Radio 
 counterparts are shown with larger filled circles, with the symbol size correlating with stellar mass. 
\textit{Top}: Rest-frame X-ray luminosity ($L_{2-10\rm{keV}}$) of DR3 sources with X-ray counterparts identified $r<2^{''}$, as a function of redshift. 
\textit{Bottom}: Rest-frame 1.4GHz luminosity ($L_{1.4\rm{GHz}}$) of DR3 sources with 3GHz counterparts within $r<0.75^{''}$. Red points indicate objects where the integrated radio emission is likely dominated by AGN activity.}
\label{fig-AGNmatch}
\end{center}
\end{figure}

We complement the DR3 photometry with the rich set of ancillary data in the COSMOS field assembled across the electromagnetic spectrum.
These will be used to constrain the dust-obscured star formation rates and investigate the prevalence of AGN. 
Appendix~\ref{sec:app-matches} provides the details of the matching between the DR3 photometric catalog and the X-ray catalog from 
\citet{civano16} and the radio catalog from \citet{smolcic17a}, as well as the analysis to estimate X-ray and radio luminosities. 
The top panel of Figure~\ref{fig-AGNmatch} shows the calculated $L_{2-10\rm{keV}}$ as a function of redshift for 
all objects in the DR3 stellar-mass complete galaxy sample 
as a function of redshift. The red curve indicates the corresponding $L_{2-10\rm{keV}}$ limit assuming the quoted X-ray source catalog depth. 
The bottom panel of Figure~\ref{fig-AGNmatch} shows the calculated $L_{1.4\rm{GHz}}$ as a 
function of redshift for all objects in the DR3 stellar-mass complete galaxy sample
as a function of redshift. Red points mark objects whose integrated radio emission 
exceeds predictions from being powered by star-formation alone (see Section~\ref{sec-agn}). 

Given the increasing importance of the dusty star-forming population amongst the most massive galaxies at early epochs \citep{martis16}, observations at far-infrared (FIR) or sub-mm wavelengths are crucial to understanding the processes that drive the rapid formation and assembly of these `monsters'. 
In order to get a better handle on the amount of star-formation occurring behind dust, we supplement the UltraVISTA DR3 (UV-IRAC) photometry of high-$z$ massive galaxies (Section~\ref{sec:DR3cat}) with cross-matched observations at MIR wavelengths and longer. 
In Appendix~\ref{sec:app-matches} we describe in detail the IR datasets used to model the full UV-to-FIR SEDs assuming energy balance 
(Section~\ref{sec-sfactivity}). Briefly, \textit{Spitzer} MIPS 24~$\mu$m fluxes were extracted, and we supplemented the UltraVISTA DR3 photometry with Herschel PACS (at 100 and 160 $\mu$m) and SPIRE (at 250, 350, and 500 $\mu$m) fluxes, SCUBA-2 850$\mu$m fluxes, and ALMA fluxes at 870 $\mu$m (Band 7) and 1.2 mm (Band 6).

We identify ($>3\sigma$) MIPS 24$\mu$m counterparts for 54 objects in the 
\textit{bona fide} high-$z$ massive galaxy sample.  
Close to half of the matched sample have low SNR 24$\mu$m fluxes (21 with $3\sigma<{\rm{SNR}_{24\mu m}}<5\sigma$). 
While this band is useful to estimate SFR obscured by dust in galaxies, the observed $24\mu$m band 
samples increasingly shorter rest-frame wavelengths at higher $z$, 
probing the rest-frame $\sim6~\mu m$ by $z\sim3$ (thermal emission from hot dust), 
arguably contaminated by emission from obscured AGN \citep{marchesini10, cowley16, alcalde19}, leading to overestimated SFRs \citep{marsan17, martis19}. 

None of the \textit{bona fide} high-$z$ massive galaxy candidates are robustly detected ($>3\sigma$) at \textit{Herschel} PACS $100~\mu$m, while only 2 are detected at 
$160~\mu$m (with ${\rm SNR}\approx4\sigma-6\sigma$). 20 candidates have $>3\sigma$ counterparts identified in \textit{at least one} of the SPIRE bands, whereas only 7 are robustly detected in \textit{all three} (250$\mu$m, 350$\mu$m and 500$\mu$m). 

We find 850$\mu$m counterparts for 26 objects in the \textit{bona fide} massive galaxy sample (4 at $z_{\rm{peak}}>4$) from the SCUBA-2 COSMOS survey (S2COSMOS; \citealt{simpson19}). Only 10 out of these 26 targets are also detected in the MIPS~$24\mu$m band (3 at $<5\sigma$ significance). 

We find ALMA counterparts for 17 galaxies in the \textit{bona fide} sample by cross-matching
with the A$^3$COSMOS dataset \citep{liu19a}.  
Three of these objects comprise of multiple components: 2 sub-mm/AzTEC sources
({`J1000+0234' and `AzTEC-5'; see \citealt{toft14}, \citealt{gomezguijarro18}) 
and the curious `Jekyll and Hyde' system at $z_{\rm{spec}}=3.717$ (see \citealt{simpson17, schreiber18b}). 

\section{Characterizing the $3<\lowercase{z}<6$ Massive Galaxy Sample}\label{sec-charpresent}

\subsection{Classification Scheme}\label{sec-dissect}

\begin{figure*}[!htbp]
\includegraphics[width=1.0\linewidth]{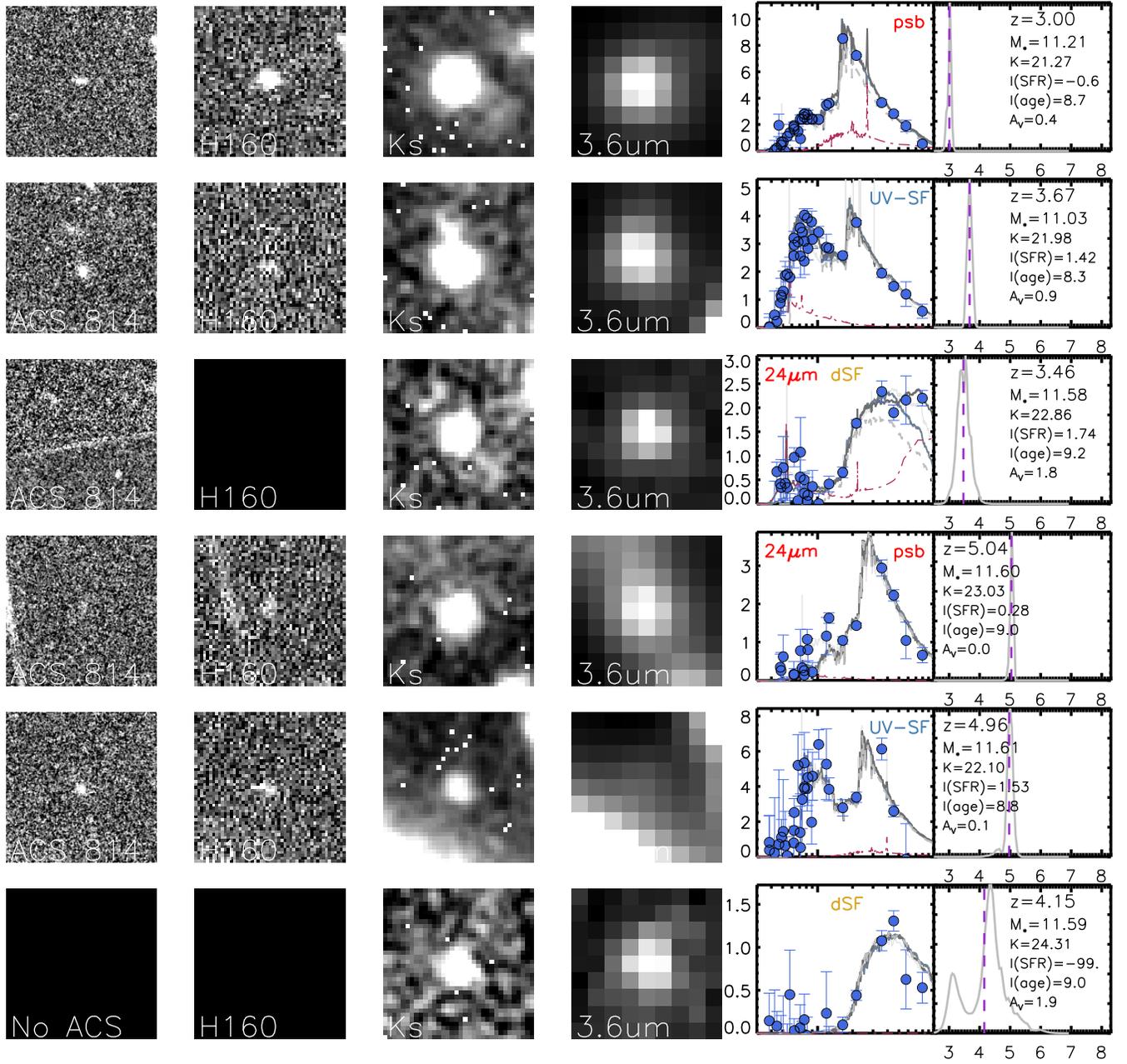}
\caption{Examples of 
massive galaxies at $3<z<4$ (first three rows) and $4<z<6$ (last three rows) sorted according to the classification scheme outlined in Section~\ref{sec-dissect}. First four columns correspond to $5''\times5''$ cutouts from available \textit{HST} ($i_{814}$ and $H_{160}$) imaging, VISTA $K_{\rm S}$ and \textit{Spitzer}~IRAC $3.6\mu$m imaging. The panels in column five displays the observed UV-IRAC photometry, along with best-fit FAST SEDs for each object (\textit{gray curves}). The best-fit AGN templates (see Section~\ref{sec-spptest:AGN}) are indicated with red dot-dashed curves. The galaxy `type' and \textit{Spitzer}~24$\mu$m $>5\sigma$ detections are listed in these panels. 
The last column shows the photometric redshift probability distribution, with the $z_{\rm{peak}}$ indicated by the vertical purple dashed lines. Best-fit (fiducial) FAST stellar population parameters are also listed. 
\label{fig-exampleUMGs}}
\end{figure*}

Given the exquisite sampling of SEDs provided by UVISTA-DR3, we decide to come up with our own method to classify the $K_{\rm S}$-selected $3<z<6$ massive galaxy population. We attempt to broadly dissect this population based on the shapes of their rest-frame UV-optical SEDs, primarily probing emission from their stellar component (see also \citealt{spitler14, forrest18}).  

Below we describe the basis on which we classified the UMGs:

\begin{itemize}
\item \textit{Dusty star-forming}: SEDs with little to- no observed UV fluxes, where primarily the rest-frame optical break is observed (redshifted to observed NIR wavelengths) with a plateau observed in the IRAC bands due to dust-heated radiation . 

\item \textit{Old and quiescent:} SEDs with little to- no observed UV fluxes, 
where primarily the rest-frame optical break is observed, although a more  ``rounded bump'' is present in the IRAC bands: \textit{we do not find strong evidence for this type of SED} 
\end{itemize}

\noindent If \textit{both} breaks are apparent with a `sharper' rest-frame optical break (i.e., \textit{concave}-like SEDs in the IRAC bands):

\begin{itemize}
\item \textit{UV-bright star-forming}: SEDs with more prominent rest-frame UV fluxes, i.e., $peak(f_{\lambda,\rm{UV}}) \gtrsim peak(f_{\lambda,\rm{opt}})$. 

\item \textit{Post-starburst}: SEDs dominated by O/A stars and 
$peak(f_{\lambda,\rm{UV}}) < peak(f_{\lambda,\rm{opt}})$.

\end{itemize}

Figure~\ref{fig-exampleUMGs} shows examples of high-$z$ massive galaxies to illustrate the galaxy classification scheme outlined above (also see Figure 6 in \citealt{forrest18}). 
This classification is consistent with the positions of galaxies on the $U-V$ vs. $V-J$ 
diagram (see Section~\ref{sec-uvj}). In the left panel of Figure~\ref{fig-histmc} we display the 
$z_{\rm{peak}}$ histogram for the \textit{bona fide} high-$z$ massive galaxy sample. 
The dusty star-forming (dSF), UV star-forming (UV-SF), and post-starburst (pSB) subsample 
are shown in gold (outlined), blue (filled) and red (hatched), respectively.   
Given the faintness of the sample, we tested for the robustness of redshift estimates using a 
Monte Carlo approach. 100 photometric catalogs were simulated 
by allowing the photometry at each observed filter to vary given the formal errors. 
The simulated catalogs were each fit with EAZY separately, and we calculated the standard deviation for the $z_{\rm{peak}}$ ($\sigma_{z}$) distribution for each galaxy. 

\begin{figure}[!htbp]
\includegraphics[width=1.0\linewidth]{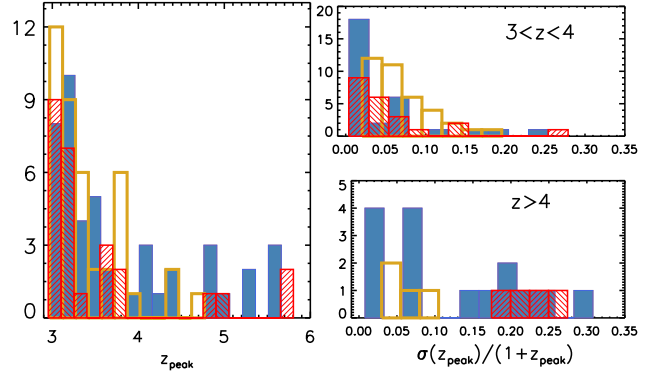}
\caption{\textit{Left panel:} $z_{\rm{peak}}$ distribution for the \textit{bona fide} high-$z$ massive galaxy sample, 
coloured according to our SED classification scheme: 
dSF - gold (outlined), SF - blue (filled) and pSB - red (hatched) histograms. 
\textit{Right} top (bottom) panel shows the distribution for the fractional error in the photometric redshifts 
($\sigma_{z}/(1+z_{\rm{peak}})$), measured by perturbing and refitting the UV-IRAC photometry of each object) 
for the \textit{bona fide} $3<z<4$ ($4<z<6$) massive galaxy sample. 
\label{fig-histmc}}
\end{figure}

The right top (bottom) panel in Figure~\ref{fig-histmc} shows the $\sigma(z_{\rm{peak}})/(1+z)$ histogram 
for the \textit{bona fide} $3<z<4$ ($4<z<6$) massive galaxy sample. 
At $3<z<4$, the $\sigma(z_{\rm{peak}})$ distribution aligns with the 1~$\sigma$ $z_{\rm{peak}}$ errors 
obtained for the fiducial EAZY runs (majority $\delta(z)<0.2$). 
The redshift probability distributions appear well constrained especially for the UV-SF and pSB population, 
as most of these objects show both the Lyman and
the optical breaks (given the young ages of the population, the latter is mostly due to the Balmer features rather than 4000{\AA} break). 
At $z>4$ however, there exist a significant fraction of massive galaxies with $\sigma(z_{\rm{peak}})/(1+z) >0.15$. These 
objects are largely undetected outside of the 
(broad and widely separated) $H$, $K_{\rm S}$ and IRAC bands, 
which prohibits noteworthy breaks in observed SEDs to be identified. 
This, combined with the rather ambiguous SED shapes of dusty galaxies 
naturally leads to the larger uncertainties in redshift estimates 
(i.e., dust and redshift being degenerate parameters in SED fitting). 
We find that although the massive galaxy $z_{\rm{phot}}$ are increasingly uncertain, they are not significant enough to affect the number of selected objects. 

\subsection{Stellar Populations}\label{sec-stellarpop}

\begin{table}[!htbp]
\begin{center}
\begin{tabular}{l  c c c  }
\hspace{1pt}  & {\textbf{pSB}} & {\textbf{UV-SF}} & {\textbf{dSF}} \\
\hline
\hline
\textbf{$3<z<4$}  & $N=29$ & $N=39$ & $N=37 $ \\ 
 & ($28\%$)  & ($37\%$)  & ($35\%$) \\
log($M_{*}/M_{\odot}$) & 11.09 (0.2) & 11.1 (0.14) & 11.2 (0.22) \\
$A_V$  & 0.6 (0.44) & 1.1 (0.37) & 2.2 (0.60) \\
log(\textit{age})$^a$ & 8.72 (0.27) & 8.86 (0.33) & 8.94 (0.45) \\
log(\textit{$\tau$})$^b$ & 7.9 (0.51) & 8.8 (0.73) & 8.70 (1.08) \\
\hline

\textbf{$4<z<6$}  & $N=4$ & $N=15$ & $N=4 $ \\ 
 & ($17\%$)  & ($65\%$)  & ($17\%$) \\
log($M_{*}/M_{\odot}$) & 11.52 (0.12) & 11.25 (0.22) & 11.38 (0.17) \\
$A_V$  & 0.15 (0.15) & 0.80 (0.46) & 1.55 (0.54) \\
log(\textit{age})$^a$ & 8.79 (0.08) & 8.82 (0.23) & 8.95 (0.08) \\
log(\textit{$\tau$})$^b$ & 7.45 (0.50) & 8.50 (0.65) & 8.60 (0.76) \\
 \\
\end{tabular}
\caption{Stellar population parameters for the \textit{bona fide} sample of massive galaxies 
at $3<z<4$ and $4<z<6$. Listed values are medians and standard deviation of the sub-sample in each redshift bin computed by 
model UV-IRAC SEDs using FAST. 
$^{a}$ denotes SFR-weighted mean stellar ages calculated for an exponentially declining SFH, where [$age$] = yr; see \citet{forsterschreiber04}. 
$^{b}$ indicates the $e$-folding star formation timescale ([$\tau$] = yr) for an exponentially declining star-formation history. }
\end{center}
\label{tab:sppchar}
\end{table}%

Table~\ref{tab:sppchar} displays the range of 
key stellar population parameters obtained for the \textit{bona fide} mass-complete galaxy sample, 
separated according to their UV-NIR SEDs in redshift bins of $3<z<4$ and $4<z<6$. 
The reported values correspond to the median and scatter calculated in each sub-sample. 
We will further explore the star-formation activity for the sample in Section~\ref{sec-sfactivity}
by incorporating constraints at longer wavelengths with MAGPHYS. We refer the reader to 
Appendix~\ref{sec-magphys} for a more detailed comparison of the best-fit stellar population parameters 
obtained using these two commonly employed SED modelling methods. Briefly, the distributions of $A_{V}$ and stellar age 
estimates are in relatively good agreement using either, however, the dusty star-forming galaxies have stellar masses 
offset to larger values ($\sim0.2$~dex) by MAGPHYS. 

Although more detailed follow-up data is required for precise analysis of star formation histories, 
given the sample size, we specifically explore trends in dust content, stellar ages and 
SFH timescales. 
The sub-samples at $3<z<4$ are characterized by similar distributions in stellar mass (log($M_{*}/M_{\odot})\approx11.1$) 
and SFR-weighted ages ($\sim0.5-0.9$~Gyrs, albeit with large scatter), implying that a large fraction of their stellar masses had formed by $z\sim5$. 
The dust extinctions and $e$-folding timescales (inferred using the fiducial exponentially declining SFH) 
however, show a range across the sample. 
As expected, the dust obscuration is largest for the dSF galaxies ($A_{V}\sim2-4$~mag), 
while the UV-SF and pSB population is characterized with $A_{V}<2$~mag. 
The pSB population is distinguished by their very short inferred star-formation timescales ($\tau<<150$~Myrs),
while the actively star-forming galaxies (both UV-bright and dust) appear to have more extended actively star-forming epochs ($\tau>400$~Myrs). 

At $4<z<6$, the median SFR-weighted stellar ages are $\gtrsim0.5$~Gyrs, 
suggesting formation redshifts as early as $z\sim7-8$. 
Similar trends are observed for dust-extinction and star-formation timescales as at 
$3<z<4$, although with much more scatter: dust extinction increases across the population (pSB to UV-SF to dSF), and 
the pSB galaxies have very short star-formation timescales ($\tau<<50$~Myrs). 

It is worth nothing that, based on their UV-NIR SEDs, and $U-V$ vs. $V-J$ colors, 
our \textit{bona fide} sample lacks a population of 
old and quiescent galaxies, which seems to rule out
quenching to not occur much before $z\approx3-5$ in the most massive galaxies at these epochs (also see \citealt{carnall20}). 

\subsection{Rest Frame $U-V$ vs $V-J$ Colors}\label{sec-uvj}

\begin{figure*}
\includegraphics[angle=90,width=1.0\linewidth]{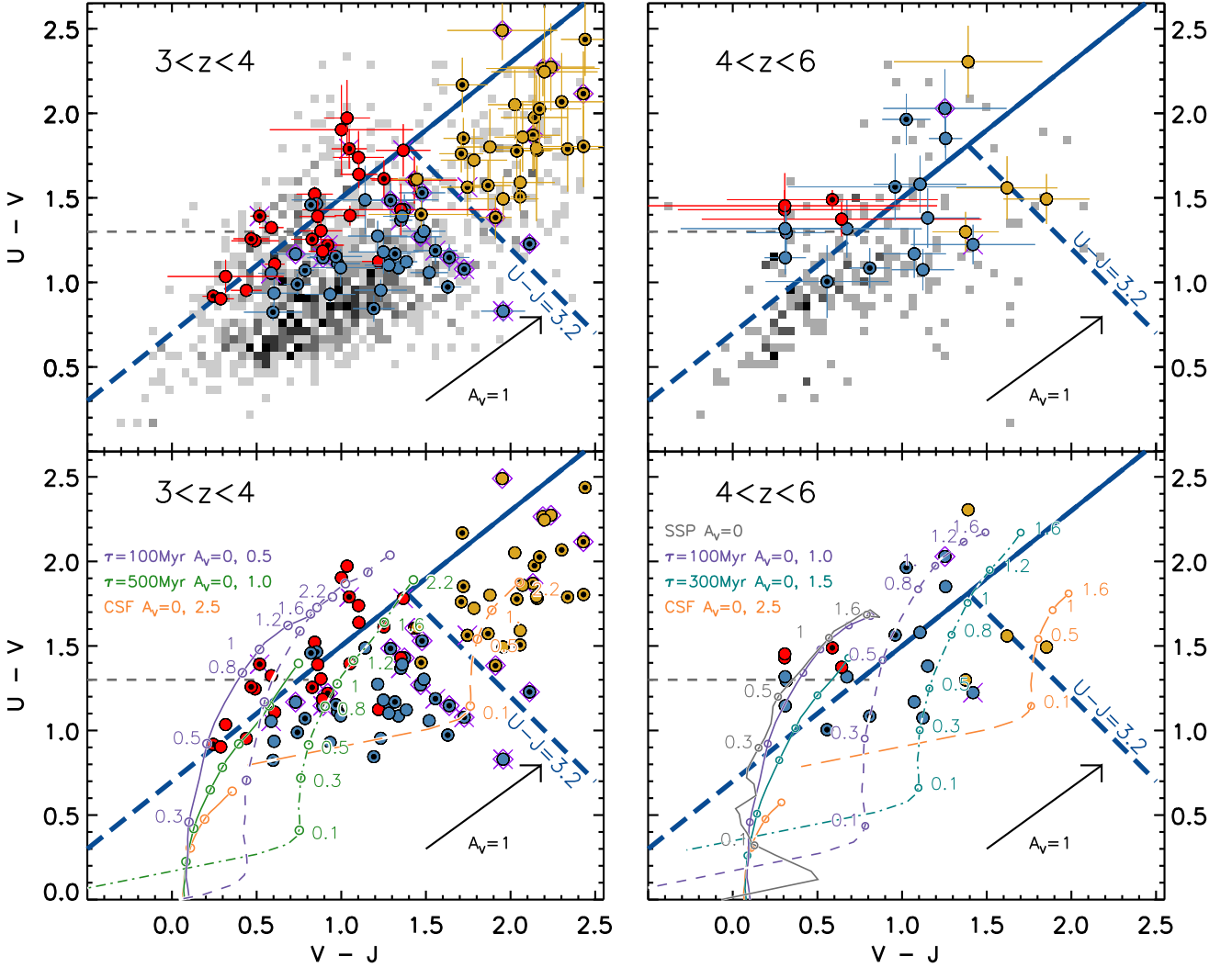}
\caption{Rest-frame $U-V$ vs $V-J$ color diagram at $3.0<z<4.0$ (left panels) and $4.0<z<6.0$ (right panels). 
The grayscale in the top panels represent the distribution of galaxies above the $95\%$ completeness limits from the $K_{\rm S}$ selected UltraVISTA DR3 catalog at the corresponding redshifts. 
\textit{Red}, \textit{blue} and \textit{gold} filled circles indicate post-starburst, UV star-forming, 
and dusty star-forming massive (log($M_{*}/M_{\odot})\gtrsim$11) galaxies, respectively. 
The thick demarcations show the extended diagonal color-color cut from \citet{whitaker15} and a slice at $U-J=3.2$ to broadly separate these three populations. 
Massive galaxies with MIPS $24\mu$m detections ($>3\sigma$) are marked with small black filled circles, while purple X and diamond symbols indicate massive galaxies with X-ray and radio counterparts, respectively. 
Errorbars (top panels) are the 1$\sigma$ distribution of the colors estimated using an MC approach. Bottom panels show color evolution tracks of \citet{bc03} models with various SFHs: 
an SSP (single burst of star formation) with no dust (gray solid curve, right panel); 
exponentially declining SFHs with 
$e-$folding timescales $\tau=100$~Myr (purple, both panels),  300~Myr (teal, right panel), and 500~Myr (green, left panel) 
with no dust (solid lines) and $A_{V}=$~0.5, 1.5, 1.0~mag (dashed curves, same colors for corresponding SFHs), respectively;  
a constant SFH with no dust (CSF; orange solid curve in both panels) and  $A_{V}=$~2.5 mag (dashed orange curves). 
The empty circles represent the model colors at the specified ages (in Gyr). The dust vector indicates an extinction of $A_{V} = 1$ mag for a \citet{calzetti00} extinction curve. 
\label{fig-UVJ}}
\end{figure*}

When large samples with deep spectroscopy are lacking, a common approach is to use 
rest-frame $U-V$ vs. $V-J$ colors \citep{williams09} to separate galaxy populations, especially at high-$z$ \citep{muzzin13b, whitaker13, straatman14}. %
Rest-frame colors were measured using the best-fit EAZY templates, as described in \citet{brammer09,muzzin13b}. Briefly, rest-frame colors are calculated from the best-fit
EAZY template. As one might expect, the robustness of calculated rest-frame colors depends on 1) how realistic of a description the chosen template set is for the specific galaxy population 2) how well sampled the galaxy SED is, at the corresponding wavelengths for the filters of color calculations. 

The top panels in Figure~\ref{fig-UVJ} display the rest-frame $U-V$ versus $V-J$ colors at $3.0 < z < 4.0$ (left) and $4.0 < z < 6.0$ (right). The grayscale represents galaxies above the $95\%$ stellar mass completeness limits in the DR3 catalog (grayscale); filled points mark the \textit{bona fide} sample 
of massive galaxies, colored according to classification established in Section~\ref{sec-dissect} (\textit{gold:} dusty-SF, \textit{blue:} UV-SF, \textit{red:} pSB). The 
rest-frame color cuts from \citet{whitaker15} are shown with the
thick solid (diagonal) and light gray dashed (at $U-V=1.3$) demarcations. 
The thick dashed lines show the color cuts we introduce to broadly separate
the population based on their SED classification:
the `quiescent wedge' is expanded to include post-starburst galaxies by
removing the $U-V>1.3$ requirement (extension shown); 
a cut at $U-J=3.2$ is used to further distinguish between dusty and relatively unobscured star-forming galaxies. The dust vector is shown for an extinction of $A_{V}= 1$~mag assuming \citet{calzetti00} extinction curve.  

The errors on the rest-frame color estimates were calculated employing a Monte Carlo approach (hence the uncertainty in redshift estimates are folded in). The standard deviation of rest-frame colors obtained in this manner for the sample are shown with the error bars in the top panels of Figure~\ref{fig-UVJ}. 
The objects in the massive galaxy sample with MIPS $24\mu$m (SNR$\geq$3) detections are indicated by small black filled circles, while purple X and diamond symbols indicate those with X-ray and radio identified counterparts, respectively.

The bottom panels in Figure~\ref{fig-UVJ} also shows color - color evolution tracks for different SFHs with varying degrees of dust extinction. 
Solid tracks represent the color evolution for SFHs assuming no dust ($A_{V} = 0$), while dashed curves indicate the dust-attenuated tracks ($A_{V}$ values labeled in panels). 
Both bottom panels include the tracks for an 
exponentially declining SFH with $\tau = 100$~Myr and $A_{V} = 0$ (solid purple curves),
a constant star formation history ($A_{V} = 0$ and 2.5~mag: orange solid and dashed curves). 
We highlight that the $\tau=100$~Myr color evolution is quite similar to the track of a SSP (solid gray track in right panel), with the latter reaching the same $U-V$ and $V-J$ colors as the $\tau=100$~Myr model
at slightly younger ages (by a few hundred Myrs). 
The bottom left panel ($3<z<4$) displays tracks for exponentially declining SFHs with 
$\tau = 100$~Myr SFH with $A_{V} = 0.5$ (dashed purple) and
$\tau = 500$~Myr ($A_{V} = 0$ and 1.0~mag: solid and dot-dashed green curves), 
while the right panel ($4<z<6$) shows tracks for 
$\tau = 300$~Myr ($A_{V} = 0$ and 1.5~mag: solid and dot-dashed teal curves). 
Small open circles along the color evolution tracks correspond to age steps since the onset of star-formation. 
The color tracks are plotted up to the maximum allowed age of the universe corresponding to the lower limit of each redshift bin, and empty circles mark model colors at the specified ages (in Gyr). 

At $3<z<4$ (Figure~\ref{fig-UVJ}, left panels), the rest-frame colors of massive galaxies span a wide range in $UVJ$ space, similar to the large spread ($>1.5$~mag) found for a mass-complete (log($M_{*}/M_{\odot}$)$~>10.6$) sample of galaxies at $3<z<4$ \citep{spitler14}. 
The massive galaxy sample at $3<z<4$ appears to be equally distributed (in numbers) in each sub-group.
The classification of massive galaxies based on their observed SEDs (outlined in \ref{sec-dissect}) aligns with expectations from the estimated rest-frame colors, and the separation between the dusty-SF (gold) and UV-SF (blue) massive galaxies is reassuringly perpendicular to the dust vector ($A_{V}$) and quiescent wedge (see also \citealt{alcalde19}).  
The dusty star-forming massive galaxies are red in both $U-V$ and $V-J$ colors, while the post-starburst massive galaxy population (red points) is located primarily towards bottom-left of the quiescent wedge, indicating young stellar ages \citep{belli19, carnall20}. 

It is not surprising that a substantial fraction of the `pSB' tagged massive galaxies at $3<z<4$ are just outside of the traditional UVJ boundary, as this criteria is known to be a relatively pure (but likely incomplete) method for selecting quiescent galaxies at high-$z$ (although see \textit{e.g.}, \citealt{martis19, roebuck19}). 
Studies have found that the horizontal cut at $U-V=1.3$ may be increasingly irrelevant at identifying passive galaxies at $z>3$, as it is possible for the colors of `dust-free' galaxies where star-formation has ceased abruptly $\sim$~few$\times100$~Myrs prior to observation to fall below this cut (\citealt{marsan15}, \citealt{merlin18}, \citealt{schreiber18}, \citealt{forrest20b}) is
common occurrence, especially at high redshifts where galaxies had little available time to evolve passively after quenching.
The colors of the `pSB' massive galaxies suggest that at $3<z<4$, observations are converging towards the epoch of quenching for UMGs. This is consistent with the picture that the typical passive massive galaxy at $z\approx3-4$ is a post-starburst, as also suggested by the analysis of \textit{HST} WFC3/IR G141 stacked grism spectrum for passive massive (log$(M_{*}/M_{\odot})\sim11.2$) galaxies at $z\sim2.8$ \citep{deugenio20}. 

Focusing on the higher-redshift bin, $4<z<6$ (Figure~\ref{fig-UVJ}, right panels), estimates of rest-frame colors are less reliable (evidenced by the larger uncertainties), as the rest-frame $V$ and $J$ bands are redshifted further into the NIR wavelengths. At $z>4$, the rest-frame $V$ band falls between $K_{\rm S}$ and IRAC $3.6\mu m$ bands, while the rest-frame $J$ band is only probed by $\approx$ IRAC $5.8\mu$m and $8.0\mu$m, where the available photometry is significantly shallower. In fact, 
all of the galaxies in our $z>4$ sample lack significant detections in IRAC $5.8\mu$m and $8.0\mu$m photometry 
($\rm{SNR}\lesssim3\sigma$). Although the full SED is taken into consideration when inferring rest-frame colors for galaxies with EaZY, at $z>4$ the photometry critical to estimating especially the rest-frame $V-J$ colors at $z>4$ are limited. Therefore we caution not to over-interpret the implied stellar populations of the red points located in the quiescent wedge.

Although we elected to choose a sample above the 90\% mass-completeness limit for UltraVISTA, 
we note that at $z>4$ our sample may be missing some highly obscured star-forming galaxies by the nature of
$K_{\rm S}$ selection \citep{caputi15, alcalde19} -- although the redshift distribution and abundance of such objects are still uncertain. 
We find that $U-V$ versus $V-J$ diagram is inadequate to constrain the stellar populations of the $4<z<6$ massive galaxy sample owing to the presence of (unconstrained) strong/high-equivalent width optical nebular emission biasing the derived rest-frame $V$ band magnitudes, along with the associated uncertainties in the derived rest-frame $J$ band magnitudes due to the low number of filters probing this spectral region at these redshifts. 

\subsection{Incidence of Active Galactic Nuclei}\label{sec-agn}

We estimate the bolometric AGN luminosity ($L_{bol, \rm{AGN}}$) for the 13 X-ray counterpart identified objects in our sample (Section~\ref{sec:xraymatch}) assuming a bolometric correction factor of $c_{bol}=10$. Following \citet{marconi04, rigby09, lusso12}, this is a conservative average for $c_{bol}$ given the parameter space in luminosity probed by the X-ray matched massive galaxies. 
Examining the top panel of Figure~\ref{fig-AGNmatch}, 
the resulting $L_{bol, \rm{AGN}}$ for the 13 X-ray matched galaxies corresponds to $L_{bol}$(AGN)$\gtrsim10^{44}$~erg$s^{-1}$. 
Only two have hardness ratio(HR) $<-0.2$, indicating that the majority of X-ray detected massive galaxies are obscured AGN. 

To identify massive galaxies whose integrated radio emission is likely dominated by AGN activity (as opposed to star-formation), we use the redshift-dependent radio excess threshold defined in \citep{delvecchio17}; $\log(\frac{L_{1.4\rm{GHz}}}{SFR_{\rm IR}})=21.984\times(1+z)^{0.013}$.  
The SFRs we use in this are those obtained with UV-FIR SED modelling with MAGPHYS (Section~\ref{sec-sfactivity}) (which does \textit{not} 
account for AGN contribution). Combining this with the fact that the majority of FIR fluxes used in the MAGPHYS modelling are merely ($Herschel$) upper limits or low-significance detections ($<3\sigma$), we argue that these values are $upper$ limits for realistic SFRs, and therefore this radio-AGN identification scheme is a conservative one. There are likely more galaxies which harbour radio-AGN than the ones we are selecting.
We find that the radio power for 8 out of the 20 radio-counterpart identified massive galaxies (Section~\ref{sec:radiomatch}) is consistent with harboring radio-AGN. These objects are indicated with red points in the bottom panel of Figure~\ref{fig-AGNmatch}. 

In summary, 19 (1) out of our sample of 105 (23) $3<z<4$ ($4<z<6$) massive galaxies show evidence of harbouring powerful AGN by the detection of luminous radio or X-ray counterparts.  
We measure the AGN fraction to be $f_{\rm{AGN}}=0.15^{+0.05}_{-0.04}$ for the full sample, while splitting the sample into two redshift bins yields $0.18^{+0.05}_{-0.04}$ and $0.04^{+0.04}_{-0.04}$ at $3<z<4$ and $4<z<6$, respectively. 
Considering instead that \textit{all} the galaxies in our sample with radio counterparts are AGN hosts 
would result in $f_{\rm{AGN}}\approx20\%$. 
We note that this estimate should be viewed as conservative lower limits since effectively 
the AGN selection methods we employed for the high-$z$ universe are biased towards brighter objects. 
These fractions are much lower than what was found in \citet{marsan17}, possibly because 
the $3<z<4$ sample here extends to much fainter objects. 

To illustrate this, we use the extended sample of galaxies with NIR spectroscopy at $3<z_{\rm{spec}}<4$ ($N=10$; whereas 5 were in \citealt{marsan17}) and repeat the analysis. 
Three objects in this sample are identified as AGN via X-ray or radio detections, while 5 are observed to have [OIII]/H$\beta$ ratios and limits consistent with ionizing source as AGN \citep{schreiber18, forrest20b}. Only 2 of the latter 5 have X-ray or radio counterparts, hence leading to 
3 additional sources to be classified as AGN by [OIII]/H$\beta$ line ratios alone (one radio-AGN identified target was found to have line ratio consistent instead with star-formation \citep{forrest20b}.
In short, considering rest-frame optical line diagnostics increases the inferred AGN fraction in the sample by a factor of 2 ($\sim30\%$ to $\sim50-60\%$). 

\section{Number Densities}\label{sec-abundances}

\subsection{Robustness of Stellar Mass Estimates}\label{sec-spptests}
Here we investigate three potential components in SED modelling which may be biasing the estimated physical parameters of our sample, in order to arrive at a `robust' 
high-$z$ massive galaxy sample with well constrained stellar masses. Specifically, we vary some assumptions made either in SED modelling using FAST (different SFH forms, inclusion of AGN) or strength of emission line contamination to photometry.

\subsubsection{Impact of Assumed Star Formation Histories}\label{sec-spptest:SFH}

\begin{figure*}
\includegraphics[angle=90,width=\linewidth]{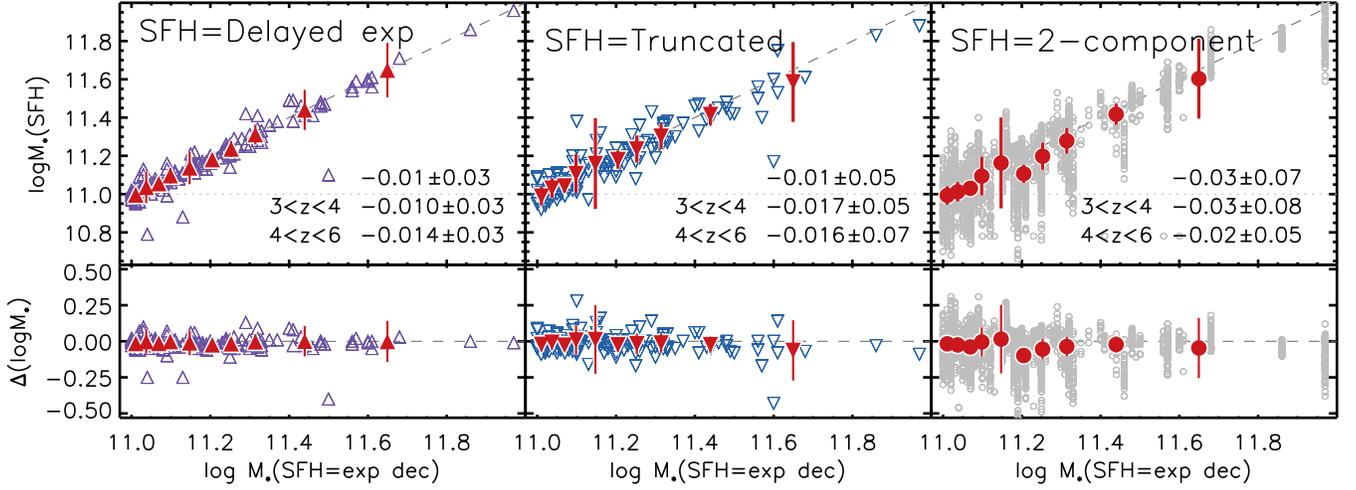}
\caption{The change in estimated stellar masses when varying the SFHs used to model UV-IRAC SEDs with FAST. 
\textit{Left}, \textit{central} and \textit{right} panels displays results when assuming the \textit{delayed exponentially declining}, \textit{constant truncated} and \textit{two-component} SFH parametrization versus the `fiducial' stellar masses used (SFH= exponentially declining). The open symbols indicate individual objects in the \textit{bona fide} $3<z<6$ massive galaxy sample. Filled red symbols and error bars mark the running means and their associated standard deviation. 
The robust mean difference in the obtained stellar masses ($\Delta{\rm log}M_{*}= {\rm log}M_{*, \rm{SFH}} - {\rm log}M_{*,\rm {SFH=exp}}$) for the assumed SFHs and their associated standard deviations are listed in the top panels (for the full \textit{bona fide} $3<z<6$ massive galaxy sample, and separated in two redshift bins).
\label{fig-dM_SFH}}
\end{figure*} 

One of the main sources of uncertainty in stellar mass estimates is the lack of knowledge of the SFHs, and it has been shown that the choice of SFH may vary stellar mass inferences by $\sim0.1-0.3$~dex \citep{michalowski14, mobasher15, carnall19a}. 
The galaxy catalog from which we selected the \textit{bona fide} sample of high-$z$ massive galaxies assumes an exponentially declining form for the SFH, which has been shown to be decent at describing the massive end of the galaxy population (provided they are not starbursts). 
However, with increasing look-back times, a larger variety of SFHs is needed to model observed SEDs. 
In this section we test the influence of assumed SFH on stellar mass estimates.

We investigated the impact of different SFHs on the estimated stellar population parameters by using two additional parametric forms for the SFH (in addition to the regular exponentially declining model): delayed exponentially declining and constant truncated SFH. 
We also consider a scenario in which a recent burst of star formation is triggered through gas infall. A recent episode of star-formation may outshine the underlying older stellar population, causing inaccurate stellar mass estimates. 
To investigate this, we created stellar libraries with \textsc{GALAXEV} \citep{bc03} assuming a two-component SFH. 
We use exponentially declining SFHs (with log($\tau$[yr]) = 7.1, 7.3, 7.5, 7.7, 7.9) to model the old stellar population accounting for the majority of a galaxy's stellar mass, and
introduce a burst contributing a fraction of the stellar mass (5\%, 10\%, 20\%, 30\%, or 50\%) when the older stellar population reaches an age of 0.2, 0.3, 0.7, 1.0, 1.2, or 1.4 Gyr. 
Using FAST, we fit the observed photometry of our sample with these different SFHs using the same grid in dust content, stellar age, and star formation timescale as in the SFH= exponentially declining case. 

Panels in Figure~\ref{fig-dM_SFH} show the resulting stellar mass estimates obtained assuming three different forms for the SFH versus the stellar masses obtained using the `fiducial' exponentially declining SFH. 
For the two-component SFH fits, given the large range in burst/old population component permutations, we impose a $\chi^2$ criterion 
to compare stellar masses that are (similarly) well-fit. 
The stellar masses corresponding to the two-component SFH fits are indicated with small gray circles in the right panel of Figure~\ref{fig-dM_SFH}. We find that applying these three SFHs 
does not cause a statistically significant offset in stellar mass estimates; stellar masses are on average 
$0.01\pm0.03$, $0.01\pm0.05$ and $0.03\pm0.07$ less than the fiducial (exponentially declining) SFH, for the 
delayed declining, truncated and the two-component SFH case, respectively.

\subsubsection{AGN Contamination to Observed UV-IRAC SED}\label{sec-spptest:AGN}

\begin{figure}
\centering
\includegraphics[width=\linewidth]{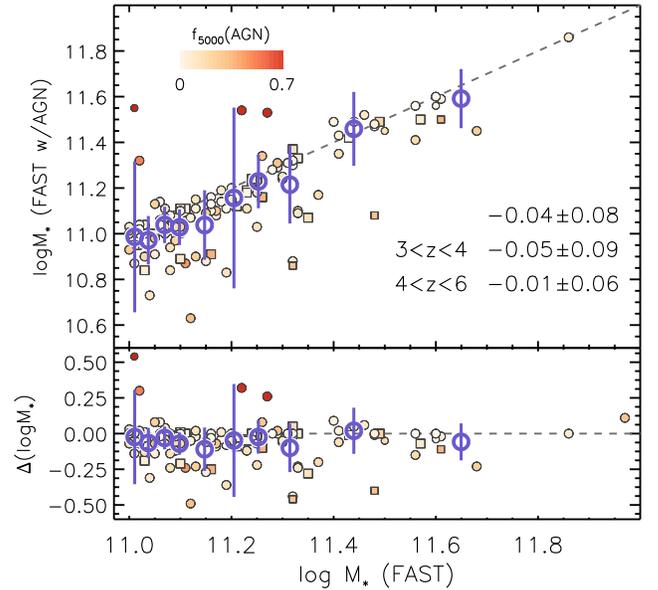}
\caption{The effect of including AGN templates in the UV-IRAC SED modelling for the \textit{bona fide} $3<z<6$ massive galaxy sample (filled symbols). 
Large open circles mark the running means and their associated standard deviation. 
The robust mean difference in the obtained stellar masses ($\Delta{\rm log}M_{*}= {\rm log}M_{*, \rm{FAST}} - {\rm log}M_{*,\rm {FAST~w/AGN}}$) and their associated standard deviations (for the full \textit{bona fide} $3<z<6$ massive galaxy sample, and separated in two redshift bins) are quoted in the top panel.  
Square symbols indicate objects that are identified as harbouring AGN in Section~\ref{sec-agn}. 
\label{fig-dM_AGN}}
\end{figure}

AGN continuum emission can potentially contribute to the rest-frame UV-optical SED, biasing the derived stellar population parameters of the sample (see \textit{e.g.,} \citealt{florez20}). 
We investigated this by using a version of the \textsc{FAST} code \citep{kriek09a} modified to perform a two component (galaxy+AGN) SED fit, presented in \citet{aird18}. This two-component fitting approach uses same grid of galaxy templates as the standard \textsc{FAST} , and a library of eight empirical AGN templates to determine the best-fit SED by marginalizing over the full grid of possible galaxy and AGN combinations. We refer the reader to \citet{aird18} for a more detailed description of this approach. Briefly, the AGN templates are comprised of five AGN-dominated SWIRE templates from \citet{polletta07} and three composite SEDs of X-ray selected AGNs from \citet{silva04}. 
We remodel the observed UV-IRAC photometry of the high-$z$ massive galaxy sample in the same manner as for the fiducial FAST runs (i.e., SFH fixed to exponentially declining parametrization and identical EAZY redshift solutions). 

Figure~\ref{fig-dM_AGN} displays the resulting stellar mass estimates when including AGN templates to the UV-IRAC SED modelling. Although it appears that the stellar masses of some massive galaxies may be biased by $>0.2$~dex, the effect on the overall population is negligible: including AGN to SED modelling in this manner decreases the stellar mass estimates on average by $0.04\pm0.08$. We stress that the derived log($M_{*}/M_{\odot}$) for individual galaxies may not be entirely accurate, as there doesn't seem to be an 
obvious offset in stellar mass estimates for the AGN-tagged massive galaxies (see Section~\ref{sec-agn}, square filled symbols in Figure~\ref{fig-dM_AGN}). In fact, the 
objects with the largest estimated AGN contribution to rest-frame optical wavelengths ($5000$\AA) appear to be massive galaxies which do not have other indications of harbouring an AGN. 
Nevertheless, this should be a sufficient sanity check for the overall sample. 

\subsubsection{Nebular Emission Line Contamination to Photometry}\label{sec-spptest:lines}
\begin{figure*}
\begin{center}
\includegraphics[angle=90,width=\linewidth]{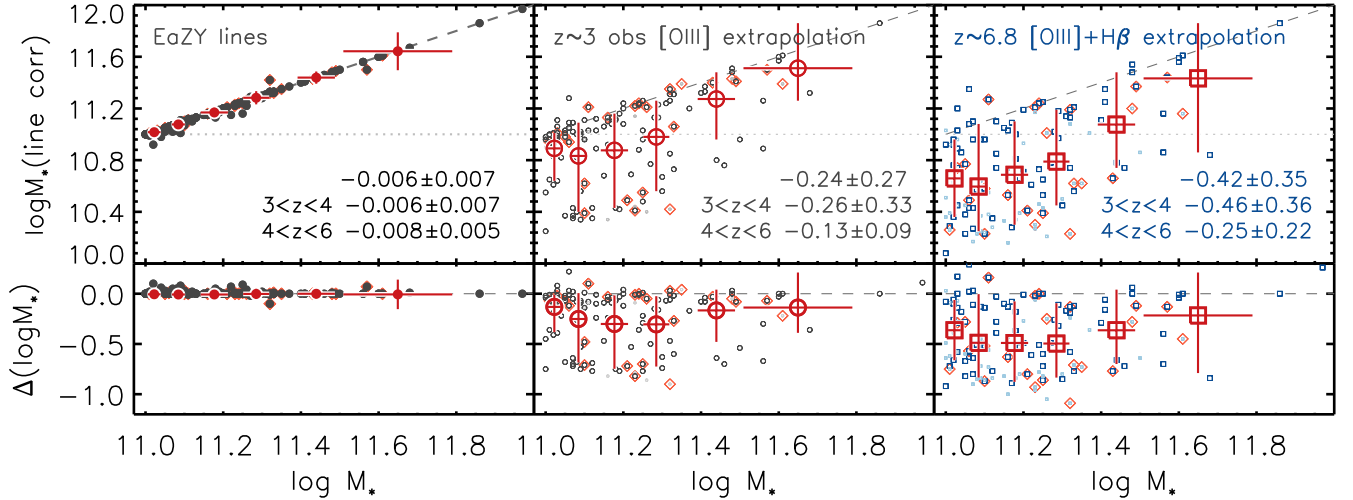}
\caption{Stellar masses obtained after correcting for nebular line contamination to photometry, using the three prescriptions for estimating equivalent widths in Section~\ref{sec-spptest:lines}. 
{\textit{Left:}} Using the EWs computed from EAzY templates.
{\textit{Center:}} Extrapolating nebular lines based on the range EWs of observed {\small{[OIII]}} emission in spectroscopic studies of $z\sim3$ massive galaxies (K-band spectroscopy). This is the more \textit{realistic} upper limit scenario for emission line contribution. {\textit{Right:}} The `extreme' upper limit to the emission line contamination prescription based on extrapolating EW ({\small{[OIII]}}+H$\beta$) estimated for galaxies at $\sim z=6.8$ by \citet{smit14}. 
Larger dark symbols indicate fits where the emission line corrected photometry yields similarly well-fit SED models as the original photometry, while small light symbols mark instances where emission-line corrected photometry is not modeled as well. Galaxies tagged as harbouring AGN are indicated with orange diamonds. The mean difference in the obtained stellar masses ($\Delta{\rm log}M_{*}= {\rm log}M_{*, \rm{orig}} - {\rm log}M_{*,\rm {line\;corr}}$) and the associated standard deviations are quoted in the top panels. \label{fig-emlinermmasses}}
\end{center}
\end{figure*}

Obtaining reliable stellar mass estimates for high-$z$ galaxies requires probing their light arising from older stellar populations (rest-frame optical or longer wavelength light).  
At $z>3$ this corresponds to observed near- and mid-IR wavelengths, where the available photometric filters are limited and may be significantly 
contaminated by emission features (due to increasing line equivalent widths with redshift), yielding overestimated stellar masses \citep{stark13, labbe13}. 
For example, a prominent/strong {\small{[OIII]}} emission feature at $3<z<4$ can mimic the NIR colors associated with 
more massive, evolved galaxies (or the Balmer break of recently quenched galaxies, see \citealt{merlin18}). 

At $z>2.5$, H$\alpha$,~{\small{[NII]}},~{\small{[SII]}} features fall out of spectral coverage range from ground, and the next most prominent optical spectral feature, {\small{[OIII]}}, is redshifted beyond the observed $K_{\rm S}$ band at $z>3.7$. Hence, current spectroscopic surveys with direct measurements of the full suite of emission lines and their equivalent widths (EW) are generally limited to the lower redshift ($z<3$) Universe (i.e., \citealt{reddy17}). At higher redshifts ($z>3$), \textit{Spitzer} IRAC photometry/colors have been employed to infer the emission-line EWs of galaxies \citep{smit14,smit16,faisst16}. 
Additionally, the strength and contribution of emission lines to IR broad-band photometry are highly uncertain at $z>4$, and will have to wait until the launch of \textit{JWST}.


The galaxy template library used to model the observed SEDs with FAST do not include prescriptions for nebular emission lines, and not accounting for 
these strong features can bias, in particular, the estimated stellar masses. 
We investigate the systematic effects that nebular line contamination to photometry may have on the derived parameters implementing the approach similar to that explored in \citet{stefanon15}; namely, by correcting each galaxy's observed photometry for potential line contamination using independent prescriptions for equivalent width estimates and recomputing the stellar population parameters in adopting the same configuration as in Section~\ref{sec:DR3cat}. The three approaches are outlined below:

\begin{itemize}
\item The \textit{minimal} correction: EW recovered from best-fit EAZY templates
\item The \textit{empirical} correction: Extrapolating the measured EW({\small{[OIII]}}) for massive galaxies at $z=3-4$ 
\item The \textit{extreme} correction: Extrapolating the estimated EW({\small{[OIII]}}+~H$\beta$) upper limits at $z\sim6.8$ 
\end{itemize}
Figure~\ref{fig-emlinermmasses} displays the effect of nebular emission contamination on the derived stellar masses assuming the three different prescriptions for equivalent width estimates outlined above.

In our first approach, we use the best-fit EAzY template, which is constructed using a combination of stellar models from the lower-$z$ universe. 
This can be thought of as the \textit{minimal} correction to the observed photometry due to emission features, because the EWs used in this approach
bias the observed photometry the least. 
This can be understood as the combination of (1)~the rising star-formation activity (on average) for galaxies with redshift (as evidenced by the evolution of the star-formation main-sequence with redshift, e.g.~\citealt{whitaker14b}) and 
(2)~lower metallicities expected for galaxies at early cosmic epochs (due to fewer cycles of chemical enrichment and on-going star-formation being fuelled by more pristine gas). 
We identified bands which could be contaminated by the (redshifted) main nebular emission lines (Ly$\alpha$, {\small{[OII]}}, H$\beta$, {\small{[OIII]}}, H$\alpha$,~{\small{[NII]}},~{\small{[SII]}}). 
The potential contribution for each feature was computed from the line equivalent width (EW) recovered from the best-fit EAzY template, and the corresponding flux was then rescaled by the factor $R(\lambda_{obs})/(\lambda_{obs}\int R(\lambda)/\lambda d\lambda)$, where $R(\lambda)$ is the filter efficiency (see Equation (1) in \citealt{smit14}).

For a \textit{realistic} emission-line correction, we adopt the maximum EW({\small{[OIII]}}) measured from $K_{\rm S}$-band spectroscopy of $z\approx3-3.5$ massive galaxies, corresponding to EW$_{rest}\approx280\pm20${\AA} \citep{marsan15,marsan17,schreiber18}. 
This value is consistent with the EW$_{rest}$ directly measured for H$\beta+${\small{[OIII]}} and {\small{[OII]}} emitters at $z\sim3.5$ \citep{khostovan15,khostovan16} in the mass range considered here. 
Lastly, for the \textit{maximal} emission-line correction to photometry, we used the EW({\small{[OIII]}}+~H$\beta$)  (EW$_{rest} \approx1550${\AA}) estimates at $z\sim 6.8$ \citep{smit14, smit15}. 
We interpret this last approach as an upper limit for nebular emission line contamination, as it is based on a lower-mass sample 
with significant excesses observed in the \textit{Spitzer} IRAC bands (which by construction require substantial EWs). 

In the latter two cases we assume a simple $(1+z)^{\beta}$ redshift evolution for EW$_{rest}$({\small{[OIII]}}), where $\beta=1.8$,  
which has been shown to describe the EW$_{rest}$ evolution for H$\alpha$ and H$\beta$+{\small{[OIII]}} (e.g. \citealt{sobral14}, \citealt{marmol16}, \citealt{rasappu16}) out to $z\sim2$, although there is evidence that a shallower ($\beta=1.3$) evolution may be present beyond $z>3$ 
(e.g., \citealt{khostovan16}, \citealt{faisst16}). 
To constrain the EW of all the (remaining) nebular emission lines adopting the line intensity ratios corresponding to $Z=0.004$ (sub-solar metallicity, $0.02 \times Z_{\odot}$) presented in Table 1 of \citet{anders03}.
For each galaxy, we computed the (estimated) observed-frame equivalent widths for each emission line considered. The necessary correction factor for each emission line was determined by comparing the estimated EW$_{obs}$ to the bandwidths of the corresponding filters that the features will fall in at the redshift of galaxies. 

The left, middle and right panels of Figure~\ref{fig-emlinermmasses} display the stellar masses obtained after correcting the observed photometry assuming \textit{minimal, empirical} and \textit{maximal} emission-line contamination to observed photometry. 
We impose a $ \Delta \chi^2 $ criterion 
to only keep values obtained by similarly good fits. We indicate the fits which satisfy $ \chi^2_{line~corr} \leq \chi^2_{orig} + 1$ by the larger and darker symbols in each panel, respectively. The average decrease in estimated stellar masses caused by the assumed \textit{minimal}, \textit{realistic} and \textit{extreme} 
emission line strength prescriptions are $0.006\pm0.007$, $0.24\pm0.27$ and $0.42\pm0.35$, respectively.

\subsection{Massive End of the Stellar Mass Function at $3<z<6$ }\label{sec:SMF}

\begin{figure*}
\includegraphics[width=\linewidth]{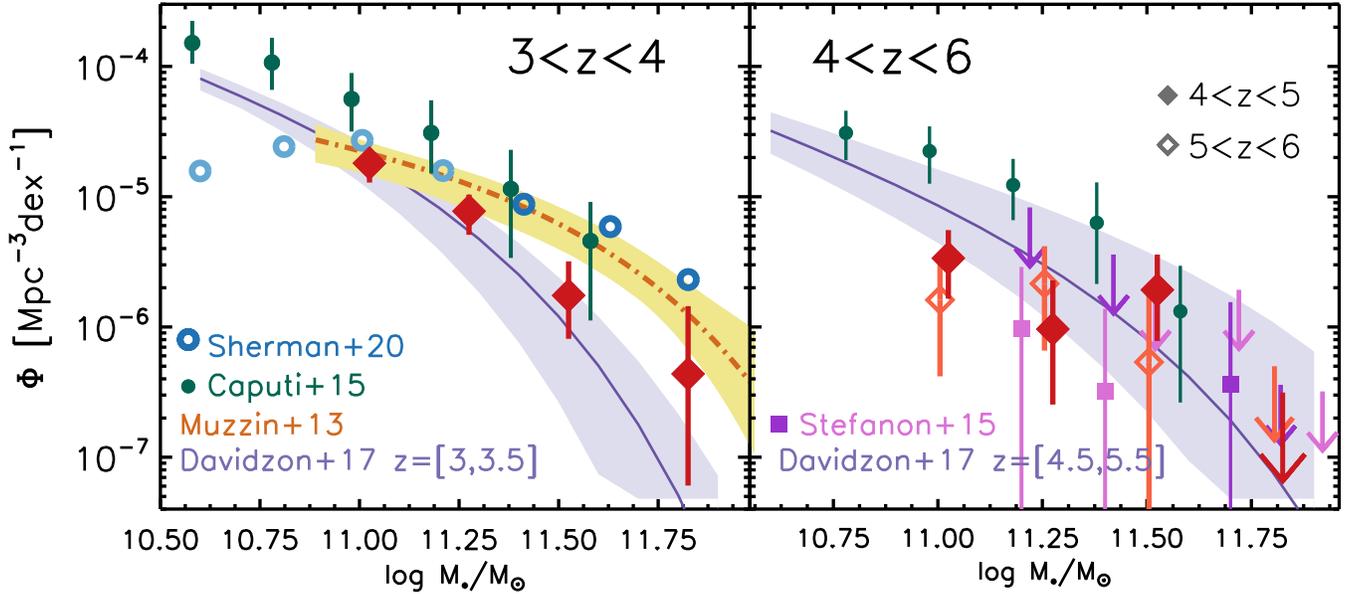}
\caption{Stellar mass functions at $3<z<4$ (left panel) and $4<z<6$ (right panel). Diamonds indicate calculations using the $1/V_{max}$ method for the \textit{robust} sample of high-$z$ massive galaxies. 
The right panel marks measurements at $4<z<5$ and $5<z<6$ with filled and open diamonds, respectively. Orange dot-dashed and purple solid curves correspond to high-$z$ SMFs ($1\sigma$ errors indicated by shaded regions) previously calculated for galaxies in the COSMOS field from \citet{muzzin13b} ($3<z<4$, left panel) and \citet{davidzon17} ($3<z<3.5$, $4.5<z<5.5$). Green filled circles show measurements from \citet{caputi15} at $3<z<4$ (left panel) and $4<z<5$ (right panel). Small blue open circles in the left panel are measurements from \citet{sherman20} for star-forming massive galaxies ($3<z<3.5$, darker symbols above their $80\%$ mass completeness limit). Pink and purple filled squares in the right panel show the SMF from \citet{stefanon15} computed at $4<z<5$ and $5<z<6$, respectively. 
\label{fig-numberden}}
\end{figure*}
 
In the previous section, 
order to infer high-confidence measurements for the abundance of massive galaxies in the early universe ~\ref{sec-spptests}.  
we demonstrated that varying the assumed SFH form (Figure~\ref{fig-dM_SFH}) or including AGN (Figure~\ref{fig-dM_AGN}) 
in SED fits does not bias the stellar mass estimates for galaxies in our sample, while the contribution of emission lines to photometry (Figure~\ref{fig-emlinermmasses})
may significantly alter their stellar mass distribution. Thus, we use the results from different emission-line correction prescriptions
to constrain the number density of very massive galaxies at $z>3$. Specifically, 
the `empirical' emission line contamination scenario is used to infer the robust number densities for the most massive galaxies at $3<z<6$, 
The \textit{bona fide} (Section~\ref{sec-sampsel}) and \textit{maximally} emission-line corrected (~\ref{sec-spptest:lines}) 
samples are used to serve as reliable upper and lower limits for the number density calculations (see Fig~\ref{fig-smf00}).

We estimate the abundance of massive (log$(M_{*}/M_{\odot})>11$) high-$z$ galaxies using the  $1/V_{max}$ formalism \citep{avni80}, 
dividing the sample into two redshift bins: $3<z<4$ and $4<z<6$, corresponding to approximately equal slices in lookback time ($\approx620$~Myrs). 
Upper and lower limits to number counts were calculated using the small sample recipe of \citet{gehrels86}. 
We estimate the contribution of cosmic variance to the total error budget using the prescription of \citet{moster11}. 
The total area covered by UltraVISTA/DR3 is $\approx0.84$~deg$^2$, 
corresponding to $\sim24\%$, $\sim35\%$, and $50\%$ fractional uncertainty at $3<z<4$, $4<z<5$, and $5<z<6$, respectively for log($M_{*}/M_{\odot}$)$\approx11.25$ galaxies. 
We assumed an additional $5\%$ fractional error on the cosmic volumes probed in the redshifts considered. The final uncertainties on density estimates were calculated by summing the above uncertainties in quadrature.

Panels in Figure~\ref{fig-numberden} display the calculated SMF at $3<z<4$ and $4<z<6$, 
diamonds indicate SMF densities computed for the \textit{robust} high-$z$ massive galaxy sample.
and include observed SMFs from literature for comparison. 
Orange dot-dashed and purple solid curves correspond to high-$z$ SMFs previously calculated for galaxies in the COSMOS field from \citet[]{muzzin13b} ($3<z<4$) and \citet[]{davidzon17} ($3<z<3.5$ and $4.5<z<5.5$), respectively. Shaded regions indicate $1\sigma$ errors. 
Green filled circles in the left (right) panel show SMF measurements from \citet[]{caputi15} at $3<z<4$ ($4<z<5$)
corrected for a \citet{chabrier03} IMF. 
In the right panel, the SMFs from \citet{stefanon15} calculated at $4<z<5$ and $5<z<6$ are plotted with 
pink and purple filled squares, respectively. 
We specifically show their `default' (i.e, no accounting for nebular emission lines) and `luminosity prior+old/dusty template' measurements. 
We note that \citet{stefanon15} computed SMFs assuming several different prescriptions for emission line corrections, the resulting SMFs are essentially all identical, consisting mostly of upper limits. 
Measurements from \citet{sherman20} reported for star forming massive galaxies identified at $3<z<3.5$ across 17.2~deg$^{2}$ are indicated with small open circles.

At $3<z<4$, our measurements are consistent with the SMF from \citet{davidzon17}, 
while lower than all other literature points. The excess seen in the SMFs from 
\citet{muzzin13b} and \citet{caputi15} can be explained by
the different survey depths used in these works, as well as the fact that 
the stellar masses inferred in these works do not account for emission line contamination
(while emission lines are included while modelling SEDs in \citealt{davidzon17}). 
The latter case especially obvious at log$(M_{*}/M_{\odot})<$11.5, where the SMF calculated using our 
\textit{bona fide} (not corrected for emission lines) massive galaxy sample is entirely consistent with these works (see Fig.~\ref{fig-smf00}). 
Interpreting the observed discrepancy with the SMF from \citet{sherman20} is less trivial, 
but can be attributed to the differences in galaxy selection methods and photometry used to construct SEDs. 
While \citet{sherman20} do implement nebular emission in their SED modelling, 
their calculations are based on measurements using (less deep, by $>2$~mag) observations with 9 photometric bands
and do not sample beyond the IRAC channel 2 band, i.e, $\lambda\gtrsim5\mu$m.

At $4<z<6$, our measurements are again below the SMF measurement from \citet{caputi15} at $4<z<5$, and 
in agreement with the SMF by \citet{davidzon17} in this redshift range. 
Our measurements are also consistent with the SMF reported in \citet{stefanon15} -- specifically, comparing 
the open (closed) diamonds indicating our measurements at $4<z<5$ ($5<z<6$) with pink (purple) squares corresponding to identical redshift intervals from \citet{stefanon15}. 

The apparently flatter SMF at $4<z<6$ is likely due to a combination of:
low number statistics, the extremely large stellar mass estimates 
(notice how in Figure~\ref{fig-masscompleteness}, 
the distribution in log($M_{*}/M_{\odot}$) is increasingly skewed towards larger values at $z>4$), and the fact that on average, stellar masses at $z>4$ appear to be less affected by contamination from nebular emission lines compared to the lower redshift bin considered (see inset text in Figure~\ref{fig-emlinermmasses}).
This means that, although the uncertainties in emission line contamination prescriptions may contribute significantly the total error budget 
in number density estimates at $3<z<4$, inferences at $4<z<6$ still appear to be dominated by low-number statistics. 

\begin{table}
\begin{center}
\begin{tabular}{| c | c c c |}
\hline
  &  & $n$~[$10^{-6}$~Mpc$^{-3}$]  &  \\ 
  \hline
  Redshift & \textit{bona fide} & Empirical Lines  & Extreme Lines \\ 
  \hline
$3<z<4$ & $10.6\pm2.7$ & $5.6\pm1.6$ & $2.4\pm0.8$ \\
$4<z<5$ & $1.7\pm0.8$ & $1.5\pm0.7$ & $1.2^{+{0.6}}_{-{0.5}}$ \\
$5<z<6$ & $1.0^{+{0.7}}_{-{0.6}} $ & $1.0^{+{0.7}}_{-{0.6}}$ & $0.7^{+{0.6}}_{-{0.5}}$ \\
\hline
\end{tabular}
\caption{Cumulative number density measurements for log($M_{*}/M_{\odot}$)$\geq11.0$ 
galaxies using the \textit{bona fide} sample and after correcting for nebular line contamination to the photometry. 
Quoted uncertainties include Poisson errors and the effect of cosmic variance. }\label{tab:numden}
\end{center}
\end{table}%

Table~\ref{tab:numden} lists the number density calculated for log($M_{*}/M_{\odot}$)$\geq11$ galaxies
for the \textit{bona fide} massive galaxy sample (Section~\ref{sec-sampsel}) and assuming the `empirical' and 
`extreme' prescriptions used to account for emission line contamination to observed photometry. 
As expected, number densities computed with the \textit{maximal} emission-line contributions are the lowest. 

At $3<z<4$, the mass-complete (${\rm{log}}(M_{*}/M_{\odot})>11$) number density we calculated for the \textit{bona fide} (emission lines not included in SED modelling) sample is 
consistent with the measurements reported in \citet{muzzin13b} ($n=9^{+6}_{-3}\times 10^{-6}$~Mpc$^{-3}$ ), as well as the mass-complete (log($M_{*}/M_{\odot})>11$) sample of galaxies identified in the CANDELS/GOODS fields 
(\citealt{alcalde19}; $14.4\pm3.7\times10^{-6}$~Mpc$^{-3}$) over the same redshift range. 
At $z>4$, the number densities calculated considering the \textit{bona fide} SED modelling scenario are lower (by $2-3\sigma$) than measurements from \citet{alcalde19} ($n=8.5\pm3.0\times 10^{-6}$~Mpc$^{-3}$ and $5.9\pm2.6\times10^{-6}$~Mpc$^{-3}$, at $4<z<5$ and $5<z<6$, respectively), with increasing discrepancy when nebular emission lines are considered to contaminate observed photometry. 
We note that the reported values from \citet{alcalde19} do not include the effect of cosmic variance in the error budget, which we estimate to be as high as $50\%$ and $110\%$ for log($M_{*}/M_{\odot})\approx11.25$ at $3<z<6$. 

In summary, accounting for nebular emission contamination to photometry causes the calculated abundances
to decrease by a factor of $\times\approx2-3$ on average, and a factor of $\sim5$ in the most extreme case. 
This highlights that ignoring the contribution of nebular emission lines in SED modelling may significantly bias the inferred properties of high-$z$ galaxies 
(although the exact magnitude of this effect is relatively unconstrained with currently available observations).

\begin{figure}
\includegraphics[width=\linewidth]{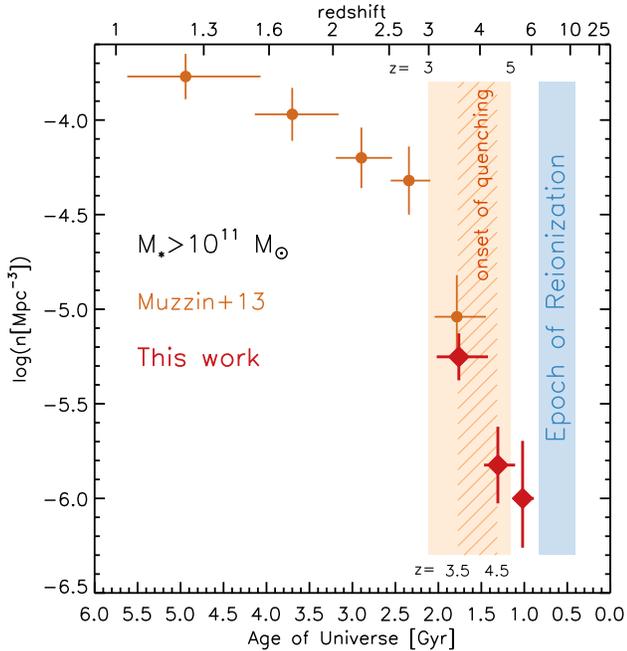}
\caption{Number density evolution for ${\rm{log}}(M_{*}/M_{\odot})>11$ galaxies 
as a function of redshift (top axis) and cosmic time (bottom axis). 
Red diamonds show our calculation assuming the `empirical' prescription for 
emission line contamination to photometry. Orange filled circles show measurements from \citep{muzzin13b}. 
The number density increases by a factor of $\gtrsim5$ and $\approx3-4$ in the $\approx 950$~Myr between $z=3-5$ (orange shaded region) and $\approx460$~Myrs between $z=3.5-4.5$ (hatched region), respectively. 
Gray shaded region represents the epoch of reionization \textbf{ ($z\approx7-11$)}. 
\label{fig-densityevol}}
\end{figure}

We show Figure~\ref{fig-densityevol} to highlight the number density evolution for ${\rm{log}}(M_{*}/M_{\odot})>11$ galaxies as a function of cosmic time (bottom axis) and redshift (top axis). 
Red diamonds indicate our measurements for the \textit{robust} sample of massive galaxies, 
while orange filled symbols show measurements from \citep{muzzin13b} to lower redshifts. 
The orange shaded region indicates the cosmic time between $z=3-5$ ($\approx 950$~Myr):
the epoch of early formation and assembly for massive galaxies, where the number density increases by $\gtrsim1.2$~dex. 
The hatched area marks $z=3.5-4.5$ ($\approx 460$~Myr), the epoch at which quenching is beginning to take place (explored further in Section~\ref{sec-sfactivity}). Finally, the blue shaded region 
shows the epoch of reionization at $z\approx7-11$ \citep{robertson15,planck18, hoag19, mason19}. 

\section{Star Formation Activity}\label{sec-sfactivity}

\begin{figure*}[!htbp]
\includegraphics[angle=90,width=\linewidth]{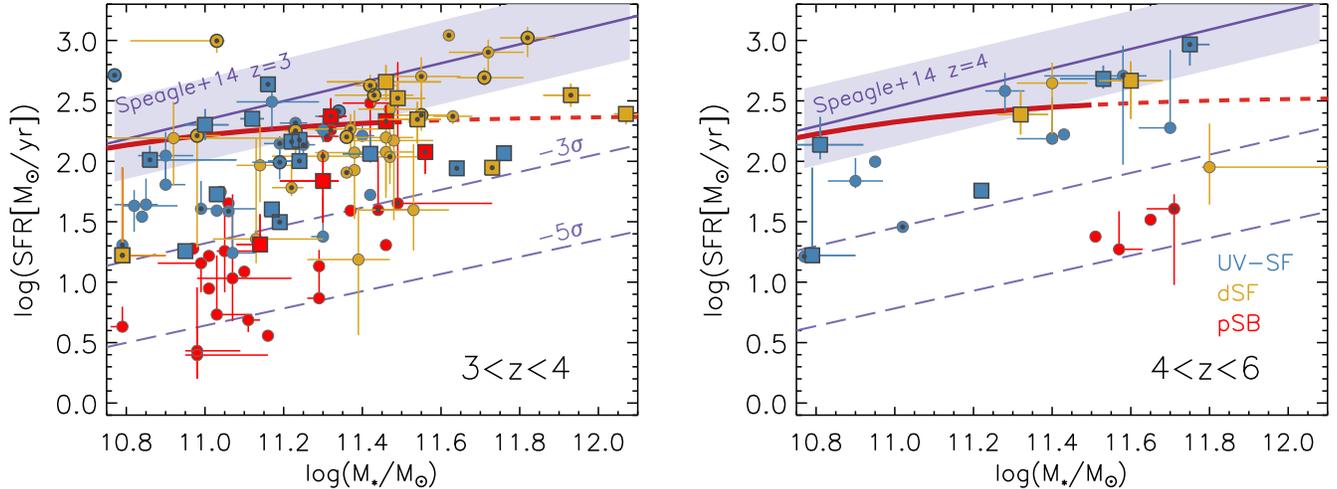}
\caption{SFRs vs. stellar mass at $3<z<4$ (left panel) and $4<z<6$ (right panel) obtained using MAGPHYS. 
\textit{Square} symbols indicate objects with FIR/radio counterparts (hence well-constrained IR SEDs), 
while \textit{filled circles} are targets lacking strong FIR constraints. 
Color-coding is identical to previous figures. 
Galaxies with $>3\sigma$ MIPS $24\mu$m, X-ray and radio counterparts are marked with 
small `bull eye' points, X's and diamonds, respectively.
Purple region shows the star-formation MS fit at $z=3$ and $z=4$ from \citet{speagle14}, with the $3\sigma$ and $5\sigma$ negative offsets are indicated with dashed lines. Red curve is the fit to the star-forming main sequence from \citet{tomczak16} at $z=3$ and $z=4$ (dashed line is extrapolation to high-mass end). 
Dotted lines indicate constant sSFR levels. 
\label{fig-sfrlogm}}
\end{figure*}

We use the counterparts identified in ancillary radio and FIR datasets (Sections~\ref{sec:radiomatch} and \ref{sec:firmatch}) to better constrain 
the amount of dust-obscured star-formation activity occurring in the \textit{bona fide} sample of massive galaxies identified in Section~\ref{sec-sampsel}.
Specifically, we incorporate the cross-matched \textit{Spitzer}~24$\mu$m, 
\textit{Herschel}~PACS ($100\mu$m, $160\mu$m)/SPIRE ($250\mu$m, $300\mu$m, $500\mu$m), 
SCUBA-2 ($850\mu$m), ALMA (Band 7/870$\mu$m, Band 6/$1.2$mm) 
and VLA (3GHz) observations to model the observed UV-radio photometry using 
the high-$z$ extension of MAGPHYS \citep{dacunha08, dacunha15}. 
This code incorporates energy balance to model the dust radiated IR SEDs, ensuring stronger constraints on the SFRs compared to estimates from UV-IRAC photometry alone. 
Although there is no AGN contribution included in the UV-FIR emission in this version of MAGPHYS, 
its fit parameters are found to be robust to AGN contamination \citep{dacunha15, hayward15}. 

Similar to FAST, MAGPHYS employs the \citet{chabrier03} IMF and the \citet{bc03} library to model stellar emission, but the two SED fitting techniques differ in several assumptions used to infer galaxy properties: 
the SFH is parametrized by a continuous delayed exponentially declining form (${\rm SFH}\propto({t~e^{-t/\tau}})/{\tau^{2}}$, 
where $\tau\equiv$~star formation timescale), superimposed with stochastic bursts of star formation; 
the two-component dust prescription of \citet{charlotfall00} is adopted to attenuate stellar emission; 
emission in the IR is modeled as described in \citet{dacunha08};
metallicity is allowed to vary from $0.2 Z_{\odot}-2 Z_{\odot}$. 
The redshifts of targets are fixed ($z_{\rm{peak}}$ from EAZY or $z_{\rm{spec}}$, when available) when running MAGPHYS. 

In Section~\ref{sec:counterparts} we identified FIR/sub-mm/radio counterparts for 38 (6) out of the 105 (23) sources in the mass-complete sample of $K_{\rm{S}}$-selected galaxies at $3<z<4$ ($4<z<6$), resulting in the extended FIR SEDs to be studied for $\sim1/3$ of the overall sample. 
For the DR3 sources lacking FIR/radio counterparts, we incorporate the upper limits on the \textit{Spitzer} and \textit{Herschel} fluxes using the depths of the corresponding surveys ($3\sigma$ upper limits of 4.5, 9.8, 9.5, 8.1, 11.4 mJy for the 100, 160, 250, 350 and 500$\mu$m bands, respectively). 
We choose to include the limits in these bands as they probe the redshifted peak of IR dust emission for $z>3$ galaxies (see for example, \citealt{schreiber18a, simpson19}), and yield more realistic constraints for SFRs compared to using more restricted UV-IRAC SED models. 
We refer the reader to Appendix~\ref{sec-magphys} for more details regarding modelling with MAGPHYS and the comparison with FAST best-fit parameters. 

Panels in Figure~\ref{fig-sfrlogm} display the median stellar mass vs. SFRs for the \textit{bona fide} sample of high-$z$ massive galaxies obtained using MAGPHYS, where SFR values correspond to the average calculated over the last 10 Myrs. 
Error bars correspond to the $16^{th}$ and $84^{th}$ percentiles calculated from likelihood distributions. 
The colors of points mark objects classification scheme outlined in Section~\ref{sec-dissect}. Square symbols indicate objects in our sample with FIR/radio counterparts 
and well-constrained FIR SEDs, while circles represent those that lack strong constraints in the FIR. 
We compare our results with the star-formation main sequence at $z=3$ and $z=4$ using the best-fit relations from 
\citet{speagle14} (purple curve, with shaded region indicating the measured scatter) and \citet{tomczak16} (red curve, dashed line indicates extrapolations to the high-mass end).

Given that the majority of the high-$z$ massive galaxy sample is relatively unconstrained in their 
IR SEDs, and that SFR estimates 
may vary widely depending on the indicator(s) used, 
we refrain from analyzing the precise 
SFR value of individual objects, but rather investigate trends in the ensemble. 
The range in SFRs is consistent with the scatter measured in other high-$z$ populations, 
although we note that extreme-starbursts appear to be missing in our sample. 
Reassuringly, galaxies that were identified as `post-starburst' based on their 
UV-IRAC SEDs are generally distributed toward lower SFRs. 
We conclude that the $K_{\rm{S}}$-selected massive galaxy sample at $3<z<6$ are largely consistent with being on the 
star-forming main sequence, with a tail extending to log(sSFR/yr$^{-1}$)~$<-10$ -- providing evidence that 
suppression of star-formation has already begun by $z\sim4$ for the most massive galaxies. 


\section{Sizes}\label{sec-size}

We take advantage of the publicly available \textit{Hubble Space Telescope} (\textit{HST}) imaging surveys in COSMOS 
to investigate the structures of high-$z$ UMGs. Specifically, we use the optical F814W ($i_{814}$, covering 1.7~deg$^2$; 
\citealt{koekemoer07}, 5$\sigma$ point source limit $i_{814}$=27.2) and NIR F160W ($H_{160}$; over 0.66~deg$^2$of the 
COSMOS field from the DASH Survey \citealt{momcheva17}, \citealt{mowla19}, 5$\sigma$ point source limit $H_{160}=25.1$) mosaics available 
over the COSMOS field. The ACS $i_{814}$ filter probes the rest-frame UV ($\approx1600-2000\AA$) wavelengths of 
galaxies at $3<z<4$, enabling a high-resolution view of star-forming regions unobscured by dust. The reddest filter on HST, 
WFC3 $H_{160}$, is better than bluer filters 
for tracing the distribution of stellar mass in high-$z$ galaxies, although we caution that $H_{160}$ is far from ideal for this, as it begins to probe rest-frame wavelengths blue-ward of the optical/Balmer break for galaxies at $z>3$. 

These mentioned \textit{HST} imaging surveys are not deep enough to detect the majority of high-$z$ massive galaxies with sufficient SNR to model their individual 2D light profiles. Therefore we opt to perform a stacking analysis to obtain constraints for the `average' size of massive galaxies at $3<z<6$. 
Two of the spec-$z$ confirmed galaxies in our massive galaxy sample have $HST/H_{160}$ band size measurements from the literature (\citealt{marsan15}, \citealt{straatman15}) - we removed these objects from the stacking analysis in the following section. 

\subsection{HST Stacking}
\begin{figure}
\includegraphics[width=\linewidth]{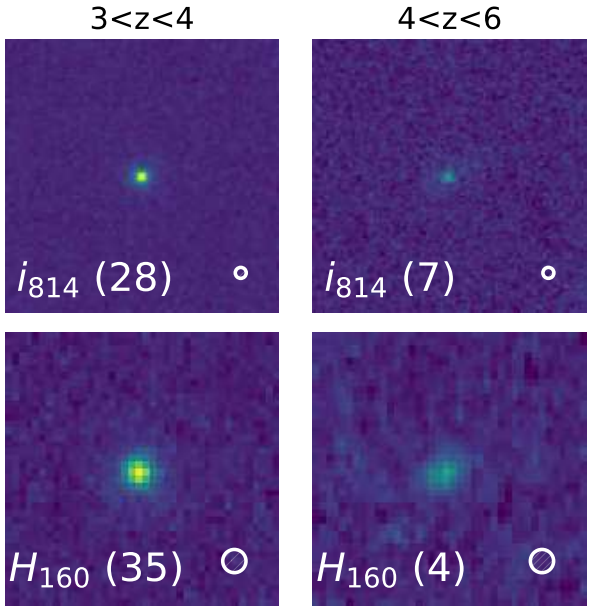}
\caption{
Stacked \textit{HST} $i_{814}$ (top row) and $H_{160}$ (bottom row) images of massive 
galaxy sample at $3<z<4$ and $4<z<6$ (left and right column, respectively). The number of objects used in each stack are quoted in parenthesis. Stamps are $4''\times4''$ in size, with 
PSFs indicated in the bottom right corner (FWHM $\approx0.''17$ and $\approx0.''08$ for $H_{160}$ and $i_{814}$ bands). 
\label{fig-HSTstack}}
\end{figure}

High-resolution stacks were constructed for the UMG sample separated in two redshift bins ($3<z<4$ and $4<z<6$) for both \textit{HST} imaging filters considered here ($i_{814}$ and $H_{160}$) in a similar manner to that presented in \citet{vandokkum10a} and \citet{hill17}. 
As a first step, we created $i_{814}$ and $H_{160}$ cutouts of size $20^{''}\times20^{''}$ centered using the coordinates for high-$z$ massive galaxies taken from the UltraVISTA DR3 catalog (corresponding to $K_{\rm S}$-band centroids). These individual cutouts were inspected to remove diffraction spike features, along with those that were contaminated by or blended with nearby bright objects. 

We constructed segmentation maps for individual \textit{HST} cutouts using \textsc{Sextractor} \citep{bertin96} {in a low SNR detection threshold setting}. 
These segmentation maps were used to identify which stamps to incorporate into our stacking (i.e., we excluded/discarded stamps that essentially had no signal associated with the location of main object in order to mitigate the effects of uncertain centroiding), as well as creating bad pixel masks which flag neighbouring objects. 
The resulting segmentation maps reveal a slight offset between the $K_{\rm S}$-band and $HST$ centroids of UMGs, and we choose to correct individual $HST$ cutouts prior to the final stacking procedure (otherwise, we will be artificially smearing the stacks). These are sub-arcsec offsets, typically $<0.''25$, and can be explained by the difference in the PSFs (VISTA $K_{\rm S}$ FWHM$\sim0.''8$ versus $HST$ FWHM$<0.''2$). This could also be due to (inhomogeneous) dust distribution since $i_{814}$ and $H_{160}$ probes ever closer to rest-frame UV wavelengths at $z>3$. 

After shifting and centering the individual \textit{HST} stamps (and their corresponding bad pixel maps), the cutouts were normalized to the observed (aperture corrected) \textit{H} band fluxes in the UltraVISTA DR3 catalog - this is to ensure that the final stacks are not biased by the brightest massive galaxies amongst the stacking sample. (\citealt{vandokkum10a} finds similar results when normalizing using either fixed aperture or aperture corrected fluxes). A \textit{final} weight map was created by summing the bad pixel masks of associated galaxies; this is used to obtain the final weighted stacked images. The final weighted stacks were created by summing the normalized and masked cutouts of individual galaxies and dividing by the corresponding \textit{final} weight map calculated in each redshift bin, for each \textit{HST} imaging filter considered here. 
Figure~\ref{fig-HSTstack} displays the $4^{''}\times4^{''}$ cutouts of the resulting \textit{HST} $i_{814}$ and $H_{160}$ image stacks for the massive galaxy sample at $z=3-4$ and $z=4-6$. Each panel lists the number of objects used to obtain the displayed stack. As expected, the stacked $HST$ images reveal the compact morphology of high-$z$ UMGs, as well as the observed dimming with increasing lookback time.

\subsection{Modelling 2D Light Profiles}\label{sec-2dlp}

\begin{table*}[htp]
\centering
\begin{threeparttable}
\begin{tabular}{|  c |  c  c  c | c  c  c  |}
\hline
 &    & $3<z<4$ &   & 
  & $4<z<6$ &   \\
\hline
Band & $n$ & $r_{e}$ [$''$] & $r_{e}$ [$kpc$]&  $n$ & $r_{e}$ [$''$] & $r_{e}$ [$kpc$] \\
\hline
{$i_{814}$}  &  $3.7\pm0.2$ & $0.089\pm0.007$  & $0.68\pm0.06$ &
\nodata & \nodata & \nodata \\
{$H_{160}$} &  $2.5, ~4$ & $0.26\pm0.01$/$0.458\pm0.01$ & $2.35\pm0.09$ &
$1,~2.5, ~4$ & $0.31\pm0.02$ &  $2.02\pm0.15$ \\
\hline
\end{tabular}
\caption{Size estimates for \text{HST} stacks of massive galaxies. Listed values are weighted averages and 1$\sigma$ uncertainty of the best-fit S\'ersic indices 
(or fixed $n$, when used in GALFIT modelling), and effect radii. }\label{tab:HSTsizes}

\end{threeparttable}

\end{table*}%

We fit the 2D light profile distributions of the high-$z$ massive galaxy \textit{HST} stacks using a single-component S\'ersic model. 
Empirical point spread function (PSF) images to be included in light profile modelling, were constructed using bright, unsaturated and isolated stars located within the $H_{160}$ and $i_{814}$ \textit{HST} mosaics. Given the compact nature of high-$z$ galaxies, measurements may be sensitive to particular PSFs. For this, and to account for PSF variations across the field, we used PSFs from 10 individual stars in each mosaic (versus stacking to obtain a single PSF). The measured PSF FWHMs are $r_{e}\approx0.''17$ and $\approx0.''08$ for $H_{160}$ and $i_{814}$ bands. 

We used GALFIT \citep{peng10} to determine the best-fit total magnitude, half-light radius ($r_{e}$), S\'ersic index ($n$), axis ratio (b/a), position angle, and the centroid for a single object. The stacked \textit{HST} images were modeled using the 10 empirical PSFs obtained from the mosaics, and the range of best-fit $r_{e}$ values were used to constrain sizes. Table~\ref{tab:HSTsizes} lists the size estimates obtained using the stacked \textit{HST} images of the massive galaxy sample. 
We convert the observed angular sizes to physical scales using the median redshift 
of galaxies in the created stacks ($i_{814}$: $\langle z \rangle=3.15\pm0.24,~4.37\pm0.35$; 
$H_{160}$: $\langle z \rangle=3.15\pm0.21,~4.88\pm0.35$). 
More details regarding the 2D light profile modelling with GALFIT can be found in Appendix~\ref{sec-galfits}. 

Figure~\ref{fig-sizemass} shows the observed stellar-mass versus size measurements for massive galaxies at $z\gtrsim3$, with our $H_{160}$ band size measurements at $3<z<4$ and $4<z<6$ shown by the green and yellow filled diamonds (error bars correspond to range of sizes obtained using fixed $n=2.5,~4.0$ fits). 
We compare with size measurements of high-$z$ galaxies from literature obtained with sub-arcsec resolution imaging. 
Blue and pink dashed lines indicate the mass-size relation at $2.5<z<3$ for star-forming and quiescent galaxies, respectively \citep{mowla19}. Dash-triple dotted lines show the extrapolation for star-forming galaxies in ZFOURGE at $z\sim3.5,~4.0,~5.0$ \citep{allen17}. Maroon and blue open circles are (UVJ-selected) quiescent and star-forming massive galaxies in ZFOURGE at 
$z\sim4$ \citep{straatman15}, while open squares represent the size measurements for quiescent massive galaxies at $z\sim4$ obtained using AO-assisted $K_{\rm S}$-band imaging \citep{kubo18}.  
Our $H_{160}$-band size measurements at both redshift bins are consistent with extrapolation of size-mass trends for star-forming galaxies at $z\sim4$ and $\sim5$ \citep{allen17}. Note that they are also just as compact as massive quiescent galaxies observed at $z\sim2.75$ \citep{mowla19}. 

\begin{figure}
\includegraphics[width=\linewidth]{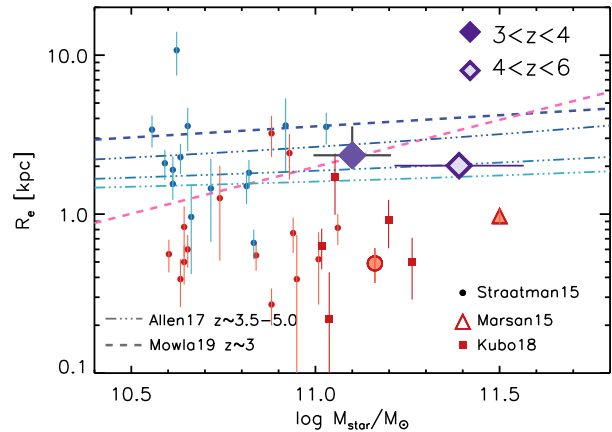}
\caption{Stellar mass and size measurements at $z\gtrsim3$. 
Filled diamonds represent effective radii obtained from the $H_{160}$ band stacks of the DR3
massive galaxy sample at $3<z<4$ and $4<z<6$. 
Filled circle and triangle symbols correspond to galaxies in our sample with available size measurements from literature (\citealt{straatman15}, \citealt{marsan15}). Other measurements from the literature are shown (see text for details). 
\label{fig-sizemass}}
\end{figure}


\section{Discussion}\label{sec:disc}

\subsection{Evolutionary Pathway to $z\sim0$}

What are these massive galaxies at $z>3$ expected to look like in the $z\sim0$ Universe? Answering this requires linking galaxy populations observed across various snapshots in cosmic time. 
Recent works investigating the progenitors of the most massive (log($M_{*}/M_{\odot}$)$>11.5$) galaxies in the local universe using abundance matching techniques \citep{marchesini14, hill17}
infer stellar mass growth by $\sim0.5$~dex since $z\sim3$ and 1.5~dex since $z\sim5$. At $z = 0.35$ galaxies with $M_{*}\approx5\times10^{11}M_{\odot}$ have a cumulative number density of $3\times10^{-6}$Mpc$^{-3}$ \citep{marchesini14}, comparable with the cumulative number density estimate for $M_{*}\gtrsim10^{11}M_{\odot}$ galaxies at $z\approx4$ (after accounting for `empirical' emission line contamination to observed photometry). 
A comprehensive progenitor/descendant analysis is beyond the scope of this work, but one can argue that  
most (if not all) $M_{*}>10^{11}M_{\odot}$ galaxies at $z>4$ must evolve to become local `ultra-massive' galaxies by simply extrapolating from previous investigations. %
These objects are likely the dense cores which accumulate stellar mass through minor mergers to become the most massive ETGs/BCGs observed at $z\sim0$.

 
\subsection{Implications for Formation Epochs and Mechanisms}

We find that the number density of massive galaxies increases by at least fivefold in the $\sim1$~Gyr of cosmic time between $z\sim3$ and $z\sim4-5$. Stacking available \textit{HST} imaging 
reveals that they are quite compact in size ($<2.5$~kpc), hinting that dissipative processes are responsible for the very early and rapid stellar mass assembly in these systems. 
Contrary to the local universe, we find that the massive end of the galaxy population at $3<z<6$ exhibits a variety of stellar populations, 
with evidence that at least some candidates display signatures of suppressed levels of star formation (either through SED shapes, or inferred SFR/sSFRs) - although the latter is less robust beyond $z>4$. 
This suggests that the processes responsible for quenching galaxies are already at play by $z\sim4$. 

Although the analysis presented in this work is based on merely photometry, 
our results corroborate previous findings 
(e.g. stellar archeological studies; \citealt{thomas05}, abundance patterns of massive 
quenched galaxies at $z\sim2$; \citealt{kriek16, belli17a, morishita19}), 
converging upon $z>5$ (and possibly as early as $z\sim7$) as the epoch at 
which the most massive galaxies in the Universe begin to form rapidly \citep{estrada20, carnall20, forrest20b}. 
We find a handful of galaxies on the way to becoming quenched at $3<z<4$, 
while the number of such viable candidates above $z > 4$ diminishes rapidly. 
This suggests that $z\approx4$ is the epoch of the first quenching of galaxies, albeit with small number statistics.
Indeed, follow-up spectroscopy of the brightest massive galaxy candidates identified in NIR extragalactic catalogs have confirmed the existence of a range of 
massive galaxies beyond $z>3$ \citep{marsan15, marsan17, schreiber18, forrest20b}, including those with no significant ongoing star formation, 
as evidenced by the absence of emission features in high SNR stellar continuum spectra out to $z\sim4$ \citep{glazebrook17, tanaka19, forrest20a,valentino20}. 

 \subsection{Limitations and Outlook}
 
It is essential to improve $z_{\rm{phot}}$ precision and accuracy (especially at $z>4$) to better constrain the observed abundances and properties 
of the most massive galaxies at high-$z$. This will also allow for targets to be efficiently selected for detailed follow-up studies. 
Increasing the sampling of SEDs is crucial for identifying evolved stellar populations 
(in particular, the rest-frame optical break feature) and improving photometric redshift estimates. NIR medium-band surveys such as NMBS \citep{whitaker11} and ZFOURGE \citep{straatman16} have 
demonstrated that observations `splitting' the $J$ and $H$ broadband filters significantly enhances the ability to identify 
$z<2$ dusty-forming galaxies which can masquerade as quiescent objects at high-$z$. 
The F2 Extragalactic Near-IR K-split (FENIKS; PI: Papovich) Survey uses split $K_{\rm{S}}$-band filters in a similar manner to enable analogous science at $z>4$ \citep{esdaile20}. 

Despite the encouraging results obtained with follow-up studies, our results show that the massive end of the galaxy population at $z>3$
extends to $K_{\rm S}=24$, revealing that the brightest objects ($K_{\rm S}<22$) currently observed are merely the `tip of the iceberg'. This highlights that there is 
much to be learned regarding the formation of the most massive galaxies (see also \citealt{forrest20b}). Extending the detailed spectral analysis (to obtain
robust stellar populations and constrain SFHs) to beyond just the brightest candidates will have to wait until the launch of \textit{JWST}. \textit{JWST} will be crucial especially at $z>4$, where most rest-frame optical features sensitive to SFH are redshifted to wavelengths not  accessible with ground based telescopes.

An obvious caveat to the $K_{\rm S}$-selection method employed (with current survey depths) is that it is effectively `blind' to the most 
dust enshrouded star-forming high-$z$ galaxies. This population is beginning to be uncovered largely thanks to ALMA, 
such as `optically-dark' galaxies \citep{schreiber18b, alcalde19, wang19} and serendipitous `ALMA-only' sources \citep{williams19, romano20, loiacono20}.

Finally, extending detailed follow-up studies to observe galaxy components beyond stars is key to understanding the formation in these remarkably early systems. 
Obtaining resolved maps of their multi-phase interstellar medium can help place meaningful limits on, for example, star-formation efficiencies and baryonic budgets. 


\section{Summary}\label{sec:summary}
 
We use the UltraVISTA DR3  $K_{\rm{S}}$-selected photometric catalog (Muzzin et al., in prep) to assemble a 
mass-complete ($M_{*}\gtrsim10^{11}M_{\odot}$) sample of galaxies at $3<z<6$, corresponding to the 
most massive galaxies in the early universe. The NIR depths and exquisite sampling of SEDs provided by the
49 band UV-IRAC photometry allow us to construct a statistical sample of high-confidence candidates 
($\approx100$ and $25$ objects at $3<z<4$ and $4<z<6$, respectively). 

The rapid number density evolution observed at $z\sim4$ (increase by a factor of $>5$ in $\lesssim500$~Myrs) 
points to the $z\sim4-6$ Universe as the epoch at which the high-mass end of the galaxy population is built up. 
The considerable fraction of the mass-complete sample showing evidence of suppressed star-formation activity ($\sim10-25\%$, depending on the preferred photometry-based diagnostic) implies that we are approaching the epoch of first quenching at $z\sim4-5$. 
The estimated stellar populations, as well as the observed trends in the rest-frame $UVJ$ colors provide additional support for this scenario:

\textbullet~At $3<z<4$, galaxies identified as post-starburst, UV-bright star-forming and dusty star-forming based on their UV-IRAC SEDs contribute equally to the massive galaxy population ($\sim30\%$ for pSB and $\sim35\%$ for both UV-SF and dSF). 
At $4<z<6$ the fraction of UV-SF identified galaxies increases by a factor of 2 (to $\sim65\%$) 
and the fraction of both pSB and dSF decreases by 2 (both to $\sim17\%$). 
SEDs of the ensemble ($3<z<6$) reveals median mass-weighted stellar ages $\sim500-900$~Myr, 
but a range in the amount of dust-obscuration and inferred SFH timescales -- consistent with the spread in rest-frame $U-V$ and $V-J$ colors. 

\textbullet~We tested the potential bias of stellar mass estimates by considering different SFHs, the inclusion of AGN templates, 
and various prescriptions for emission-line contamination to the observed photometry in UV-IRAC SED modelling. 
We measure robust abundances for the most massive galaxies (log$(M_{*}/M_{\odot})>11$) at $3<z<6$ and present the updated SMFs. 
Our findings are largely in agreement with previous measurements in literature; 
however, the improved statistics made available by deeper NIR measurements allow us to highlight the rapid emergence of 
the most massive galaxies at $z\approx4$. 

\textbullet~
We incorporate ancillary data at FIR-radio wavelengths to model the extended UV-FIR SEDs of the mass-complete sample using MAGPHYS with the aim of 
obtaining reliable constraints for their IR luminosities and dust-obscured star-formation rates. Despite their extreme stellar masses, the (UV-bright and dusty) 
actively star-forming population is consistent with being on the star-formation main-sequence at $3<z<6$. 
There is a large range in inferred SFRs and they are in general agreement with our classification based on UV-IRAC SEDs: $\sim10-15\%$ ($\sim25\%$) of
the selected sample at $3<z<4$ ($4<z<6$) have $SFR\gtrsim300$~M$_{\odot}/yr$; at both redshift bins, $\sim15-20\%$ of the population show evidence of suppressed star-formation activity ($<3\sigma$, i.e., $\approx 1$~dex below MS, or $sSFR \lesssim 0.1$~Gyr$^{-1}$). 

\textbullet~We perform stacking analysis using available \textit{HST} imaging across the COSMOS field ($i_{814}$ and $H_{160}$ bands) 
to obtain size estimates for the sample at $3<z<4$ and $4<z<6$. Consistent with previous works, we find the massive galaxy population at high-$z$ is remarkably compact ($r_{e}(H_{160})\lesssim 2.3$~kpc and $r_{e}(i_{814})\lesssim 0.6$~kpc). 

\acknowledgements
Z.C.M. gratefully acknowledges support from the Faculty of Science at York University as a York Science Fellow.
This work was supported by the National Science Foundation through grants AST-1513473, AST-1517863, AST-1518257 and AST-1815475
by \textit{HST} program number GO-15294, and by grant numbers NNX16AN49G and 80NSSC17K0019 issued through the NASA Astrophysics Data Analysis Program (ADAP). Support for program number GO-15294 was provided by NASA through a grant from the Space Telescope Science Institute, which is operated by the Association of Universities for Research in Astronomy, Incorporated, under NASA contract NAS5-26555. 

Based on data products from observations made with ESO Telescopes at the La Silla Paranal Observatory under ESO programme ID 179.A-2005 and on data products produced by CALET and the Cambridge Astronomy Survey Unit on behalf of the UltraVISTA consortium.


\appendix


\section{Multi-Wavelength Counterpart Identification}\label{sec:app-matches}
\renewcommand\thefigure{\thesection\arabic{figure}}   
\renewcommand\thetable{\thesection\arabic{table}}    

\setcounter{figure}{0}
\setcounter{table}{0}

\subsection{X-ray Counterparts}\label{sec:xraymatch}
We use the publicly available Chandra COSMOS Legacy source catalog (\citealt{civano16}, limiting depth $8.9\times10^{-16}$~erg~cm$^{-2}$~s$^{-1}$ 
in the full 0.5-10~keV band) to search for X-ray counterparts (with \textrm{DET\_LM} $\geq10.8$) around the positions of DR3 $K_{\rm S}$-band source catalog 
using a search radius of $r=2''$. 
We convert the observed full-band X-ray fluxes to rest-frame 2-10~keV luminosities ($L_{2-10\rm{keV}}$) for sources according to 
Equation (4) in \citet{marchesi16} assuming a fixed photon index of $\Gamma=1.4$, and fixing their redshifts to best-fit $z_{\rm{phot}}$ solutions, or spectroscopic redshifts, when available. 

\subsection{Radio Counterparts}\label{sec:radiomatch}
We utilize the publicly available source catalog from the VLA-COSMOS 3~GHz Large Project (\citealt{smolcic17a}, reaching median rms 
$\sim2.3\mu$Jy/beam at an angular resolution of $0.75''$) to search for radio counterparts of UltraVISTA DR3 sources. 
We identify radio-counterparts for 22/128 of objects in the \textit{bona fide} high-$z$ massive galaxy sample, 
using a search radius of $r=0.7"$ and a $SNR>5.5$ cut (to account for spurious sources 
as recommended in \citealt{smolcic17a}). 
Assuming a radio spectral index $\alpha=-0.7$ (i.e., $S_{\nu}\propto\nu^{\alpha}$) 
and the best-fit EAZY (or spec-$z$ when available) redshifts to convert the observed-frame 3~GHz flux densities to rest-frame 
1.4~GHz luminosities. 

\subsection{Infrared Counterparts}\label{sec:firmatch}


\textbullet~\textbf{Spitzer MIPS 24$\mu$m Counterparts:} 
The $24\mu$m photometry for objects in the DR3 catalog were measured on deblended images 
constructed using the positions of $K_{\rm S}$ band detections. Briefly, 
all sources in the $K_{\rm S}$ band were convolved with a kernel derived 
from bright PSF stars in the $K_{\rm S}$ and \textit{Spitzer} MIPS 24$\mu$m 
images \citep{lefloch09}. An individual ``cleaned'' $24\mu$m image 
was produced for each object after subtracting off the total flux of all nearby sources, 
and the $24\mu$m flux for each object was obtained by performing aperture photometry on these images 
($5''$ diameter aperture).  
We refer the reader to \citet{wuyts08} for a more detailed discussion and demonstration of this method.

\textbullet~ \textbf{{$Herschel$ PACS/SPIRE Counterparts:}} 
We supplement the UltraVISTA DR3 photometry using the publicly available \textit{Herschel} data from the 
\textit{Herschel} PACS Evolutionary Probe (PEP; \citealt{lutz11}),
 \textit{Herschel} Multi-Tiered Extragalactic Survey (HerMES; \citealt{oliver12, hurley16}, with SPIRE 250$\mu$m, 350$\mu$m and 500$\mu$m filters). 
The $3\sigma$ depths for PACS imaging at 100$\mu$m and 160$\mu$m are 4.5 and 9.8~mJy ($\rm{FWHM}=7''-11''$), 
while for the SPIRE 250$\mu$m, 350$\mu$m and 500$\mu$m maps they are 9.5, 8.1, and 11.4 mJy ($\rm{FWHM}=18''-35''$). 
For both surveys, we use the catalogues of fluxes extracted using Spitzer $24\mu m$ prior positions.
We match sources with at least 3$\sigma$ detection in any of the Herschel bands to MIPS-detected sources in the UltraVISTA DR3 
catalog using a matching radius of $1.5''$. 
We refer the reader to \citet{martis19} for a more detailed discussion of matching and flux extraction technique.

\textbullet~\textbf{SCUBA-2~850$\mu$m Counterparts: }
We cross-match our sample of $K_{\rm S}$-selected $3<z<6$ massive galaxy sample with publicly available $850\mu$m source catalogue 
provided across the COSMOS field by the SCUBA-2 COSMOS survey (S2COSMOS; \citealt{simpson19}). The source catalogue
consists of $850\mu$m sources detected at a significance of $\gtrsim4\sigma$ across $\sim2.6$~square degrees in the COSMOS field (median noise $\sigma = 1.2-1.7$~mJy~beam$^{-1}$, $\rm{FWHM} = 14''.9$). 
We use a search radius of $r\lesssim3''$ for counterpart identification. 
The $S_{850\mu m}$ flux densities are in the range 
$\approx2.0-13.2$~mJy with a median measured value $\sim5.5$~mJy. 

\textbullet~\textbf{ALMA Counterparts:} 
Wide-field FIR/sub-mm surveys conducted using single-dish instruments generally lack the sensitivity to place meaningful constraints on the star formation in sources beyond the most IR-luminous high-$z$ galaxies.
Their large beam sizes (and hence, low angular resolution) also mean that their analysis is riddled with source-blending issues (e.g., \citealt{karim13, trakhtenbrot17, wardlow18, simpsonj20}). 
To capture the FIR SEDs for selected objects with higher angular resolution and deeper flux limits,
we augment the photometry of massive high-$z$ galaxies taking advantage of the A$^3$COSMOS \citep{liu19a} dataset constructed using publicly available Atacama Large Millimeter Array (ALMA) archival data in the COSMOS field. 
Specifically, we use the blind-extracted source catalog (SNR$_{peak}>5.4$ in Band 6 and/or 7 continuum data) using a search radius of $r=1"$, finding matches for 17 of the high-$z$ massive galaxies (2 at $z_{\rm{phot}}>4$). 
We note that although this dataset overcomes source-blending issues, it is based on a compilation of various pointed surveys, leading to mixed coverage and depths across the field. 
There are 5 targets with Band 7 (870$\mu$m) flux densities in the range 
$S_{\nu}\sim1.5-9$~mJy (median $S_{870}\approx4.7$~mJy), 
while 16 have Band 6 ($\sim$1.2~mm) measurements in the range 
$S_{\nu}\sim0.3-3.6$~mJy (median $S_{1.2}\approx1.8$~mJy).

\renewcommand\thefigure{\thesection\arabic{figure}}   
\renewcommand\thetable{\thesection\arabic{table}}    
 
\setcounter{figure}{0}
\setcounter{table}{0}

\section{Calculated SMFs with Different Emission-Line Contamination Prescriptions}

\begin{figure*}[!htbp]
\includegraphics[width=\linewidth]{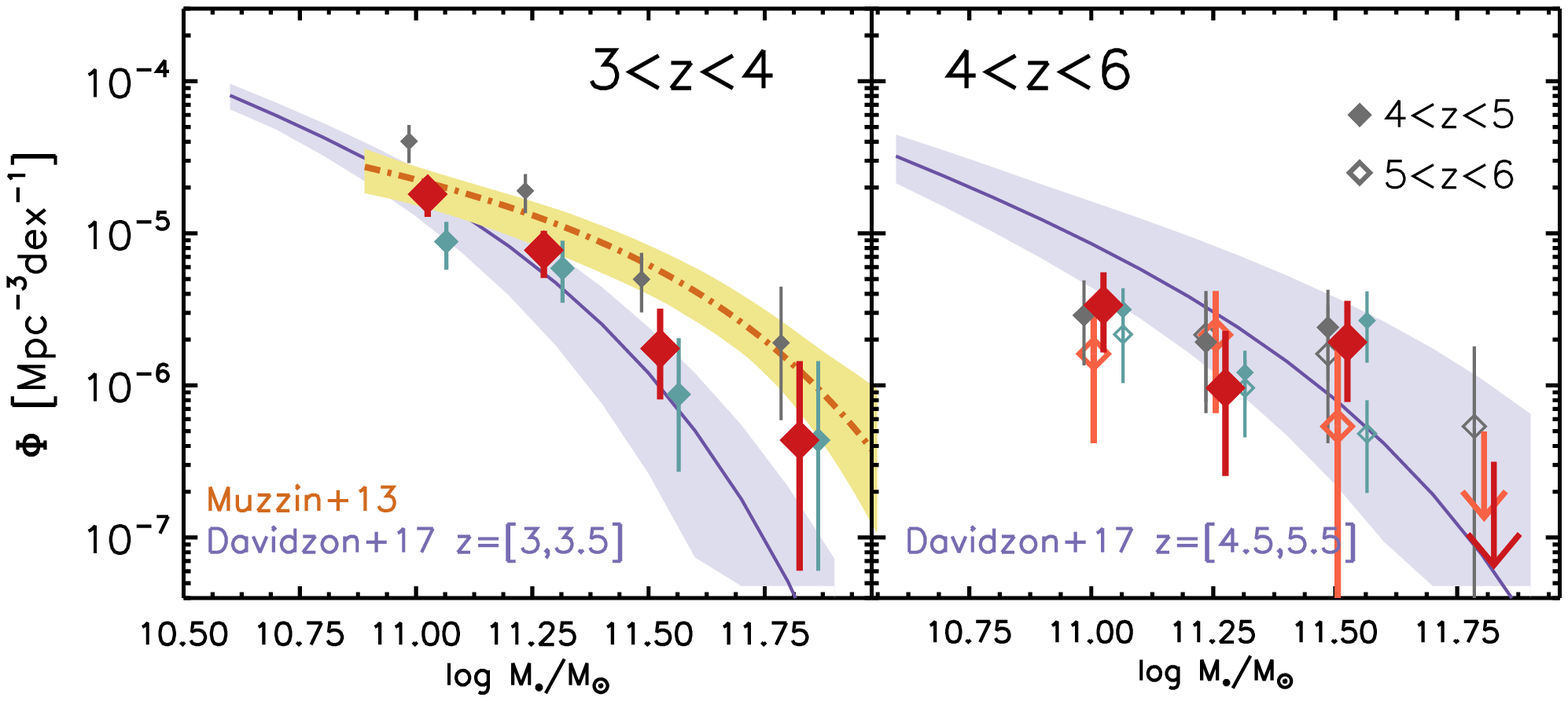}
\caption{Stellar mass functions at $3<z<4$ (left panel) and $4<z<6$ (right panel). Diamonds indicate calculations using the $1/V_{max}$ method for the robust (large, red), \textit{bona fide} (medium, gray) and \textit{maximally} emission-line corrected (small, light blue) sample of high-$z$ massive galaxies. Shaded regions show, for comparison, the SMFs from \citet{muzzin13b} and \citet{davidzon17}, same as plotted in Figure~\ref{fig-numberden}.
\label{fig-smf00}}
\end{figure*}

Panels in Figure~\ref{fig-smf00} display the SMFs calculated at $3<z<4$ and $4<z<6$ using the three approaches to estimate nebular emission line contamination to the observed UV-NIR photometry (Section~\ref{sec-spptest:lines}). 
Red (large) diamonds in Figure~\ref{fig-smf00} indicate the densities computed for the robust high-$z$ massive galaxy sample (same as in Figure~\ref{fig-numberden}). The same calculations, repeated using the \textit{bona fide} sample and the sample corrected assuming the `extreme' emission line prescription
are plotted with gray and light blue diamonds, respectively. 

It is evident from Figure~\ref{fig-smf00} that the 
calculated SMFs for massive galaxies at $3<z<4$ depend heavily on the assumed emission-line prescriptions, 
while low-sample statistics still dominates the total error budget at $4<z<6$. 
 
\section{Comparing Stellar Populations Estimated by FAST and MAGPHYS}\label{sec-magphys}
\renewcommand\thefigure{\thesection\arabic{figure}}   
\renewcommand\thetable{\thesection\arabic{table}}    
 
\setcounter{figure}{0}
\setcounter{table}{0}

\begin{table}
\begin{center}
\begin{tabular}{l  c c c  }
\hspace{1pt}  & {\textbf{pSB}} & {\textbf{UV-SF}} & {\textbf{dSF}} \\
\hline
\hline
\textbf{$3<z<4$}  & $N=29$ & $N=39$ & $N=37 $ \\ 
log($M_{*}/M_{\odot}$) & 11.15 (0.20) & 11.17 (0.24) & 11.46 (0.27) \\
$A_V$  & 0.73 (0.58) & 0.84 (0.59) & 2.90 (0.75) \\
log(\textit{age}) & 8.85 (0.17) & 8.96 (0.21) & 8.91 (0.23) \\
log(\textit{$SFR$})$^a$ & 1.27 (0.61) & 2.01 (0.38) & 2.23 (0.43) \\
\hline
\textbf{$4<z<6$}  & $N=4$ & $N=15$ & $N=4 $ \\ 
log($M_{*}/M_{\odot}$) & 11.61 (0.08) & 11.28 (0.33) & 11.5 (0.19) \\
$A_V$  & 0.94 (0.27) & 0.76 (0.70) & 2.41 (0.30) \\
log(\textit{age}) & 8.76 (0.07) & 8.80 (0.13) & 8.76 (0.08) \\
log(\textit{$SFR$})$^a$ & 1.45 (0.13) & 2.19 (0.51) & 2.52 (0.28) \\
 \\
\end{tabular}
\caption{Stellar population parameters for massive galaxies 
at $3<z<4$ and $4<z<6$. Listed values are medians and standard deviation of the sub-sample in each redshift bin computed by 
model UV-FIR SEDs using MAGPHYS. $^a$: $SFR$ in units of $M_{\odot}/yr$. }
\end{center}
\label{tab:sppchar-magphys}
\end{table}%

Here we compare the best-fit stellar population parameters obtained for the \textit{bona fide} massive galaxy sample using FAST and EaZY. As the massive galaxy population becomes increasingly more dominated by dust obscured star-formation, we tested how the estimated stellar population properties varied when FIR constraints are considered in SED fitting. 
We specifically considered the key stellar population parameters investigated in this work:
log($M_{*}$), SFR, $A_{V}$ and stellar age. 
Here we remind the reader that (more restricted) UV-NIR SEDs are used with FAST (Sections~\ref{sec:DR3cat}, \ref{sec-spptests}) versus the extended UV-FIR/radio SEDs with MAGPHYS (Section~\ref{sec-sfactivity}). 
\citet{martis19} report that $M_{*}$, $A_{V}$, and SFR estimates obtained using MAGPHYS are fairly robust to \textit{Herschel} photometry. 

Table~\ref{tab:sppchar-magphys} lists the average log($M_{*}$), log(SFR), $A_{V}$ and log(age) 
MAGPHYS estimates for the sample, separated according to the classification in Section~\ref{sec-dissect}. 
Panels in Figure~\ref{fig-fastvsmagphys} show the comparison of $M_{*}$, $A_{V}$, and SFR and stellar age estimates for the high-$z$ massive galaxy sample (defined in Section~\ref{sec-charpresent}) using FAST and MAGPHYS. 
We use the median ($50^{th}$) and the corresponding $16^{th}$ and $84^{th}$ percentiles for the parameter distribution from MAGPHYS. 
The distributions for best-fit parameters obtained with these two methods agree relatively well, albeit with some inconsistencies, which we outline below.

The estimated stellar masses for the post-starburst (red symbols) and UV-SF (blue symbols) population are consistent within $\sim\pm0.2$~dex, comparable with the
variation reported in many previous works, e.g., \citet{michalowski14, mobasher15, carnall19a}. 
The MAGPHYS stellar mass estimates of the dusty-SF galaxies, however, are offset by $\sim+0.25$~dex compared to those calculated by FAST (see also \citealt{liu19a}). 
The largest discrepancy between the investigated stellar population parameters is seen in the \textit{SFR} estimates (top right panel in Figure~\ref{fig-fastvsmagphys}). This is expected, as the implementation of FIR constraints to account for dust re-emission fundamentally alters how the two method calculate SFRs. FAST underestimates the total SFR 
as by construction it does not know anything about fluxes beyond $\lambda_{\rm{obs}}>10\mu$m -- where dust re-emitted radiation is significant.  

\begin{figure}
\includegraphics[width=\linewidth]{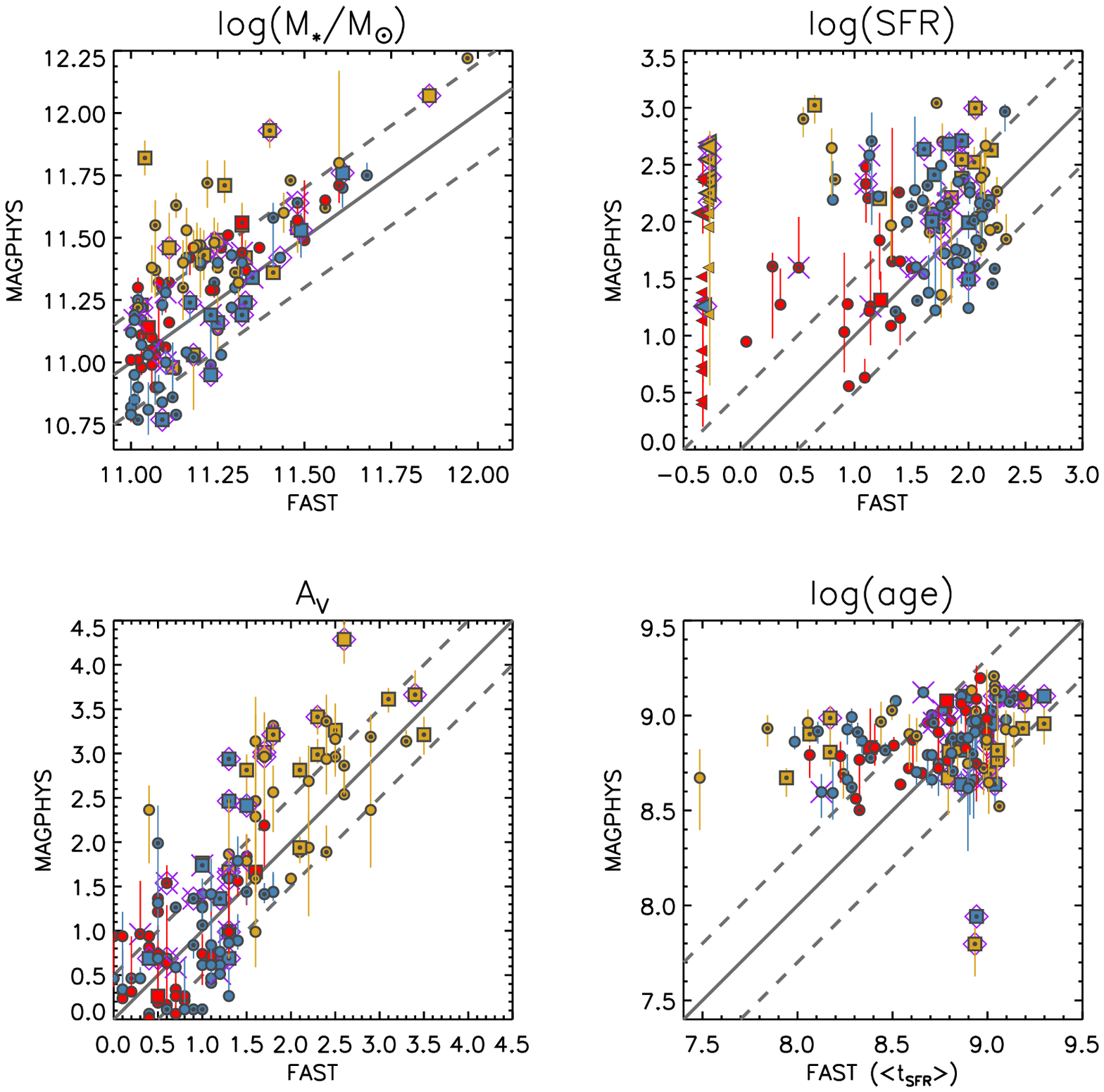}
\caption{Comparison of best-fit stellar-population parameters obtained by modelling UV-NIR SEDs with FAST versus the modelling with MAGPHYS, which incorporates \textit{Spitzer} $24\mu$m, \textit{Herschel} (PACS and SPIRE), SCUBA-2 $850\mu$m, ALMA (Band 6 and 7) detections and upper limits. 
Symbols are color-coded according to the dissection of the massive galaxy sample in Section~\ref{sec-dissect}. Solid diagonal lines show 1:1 relation, dashed lines indicate: $\pm0.25~dex$ in log(M$_{*}$),  $\pm0.5~dex$ in log($SFR$), 
$\pm0.5$~mag in $A_{V}$, and $\pm0.3~dex$ in log($t_{\rm{age}}$). 
\label{fig-fastvsmagphys}}
\end{figure}

\section{2D Light Profile modelling of HST Stacks}\label{sec-galfits}
\renewcommand\thefigure{\thesection\arabic{figure}}   
\renewcommand\thetable{\thesection\arabic{table}}    
 
\setcounter{figure}{0}
\setcounter{table}{0}


For the $i_{814}$ $3<z<4$ stacks, the best-fit S\'ersic indices and half-light radii ranged between $n\sim3-6$ and $r_{e}\sim2.3-3.4$~pixels (corresponding to $\sim0.''07-0.''1$, indicating that it is just barely resolved). 
We were not able to obtain reliable size measurements for the high redshift ($4<z<6$) $i_{814}$ stacks given the low $SNR$. Regardless, Figure~\ref{fig-HSTstack} shows that they appear to be as compact as their $3<z<4$ counterparts. 
Interpretation of size estimates based on $i_{814}$ band images are complicated due to dust obscuration considerations.

Modelling the light profile of the $H_{160}$ image stacks did not converge when all the S\'ersic profile parameters were allowed to vary. Therefore, size estimates were obtained using fixed $n=1.0$,~2.5,~4.0. 
For the $3<z<4$ $H_{160}$ stack modelling, the $r_{e}$ best-fit estimates were correlated with the
specific $n$ S\'ersic profiles used: $\langle r_{e} \rangle = 0.''175\pm0.''003$, $0.''260\pm0.004''$, $0.''458\pm0.008''$, for $n=1$,~2.5,~4.0, respectively (weighted averages of fits using the 10 different PSFs). 
Visually inspecting the residuals from the $n=1$ models fits showed them to be not good fits (over-subtracting a compact center); we therefore do not consider them further. 
The $4<z<6$ $H_{160}$ $r_{e}$ estimates obtained for all the fixed $n$ profiles were consistent with each other (within 1$\sigma$ uncertainties). We calculate the weighted averages sizes for the $3<z<4$ and 
$4<z<6$ stacks of $H_{160}$ imaging to be $\langle r_{e} \rangle = 0.''302\pm0.''012$ and $0.''310\pm0.''023$.

Figure~\ref{fig-galfits} displays examples of single-component S\'ersic profile models for the \textit{HST} stacks analyzed in Section~\ref{sec-2dlp}.
Columns are: \textit{HST} stack image cutouts, best-fit 2D light profile obtained with GALFIT ($n$ and $r_{e}$ indicated in row) and the residual image. 

\begin{figure}
\begin{center}
\includegraphics[width=0.6\linewidth]{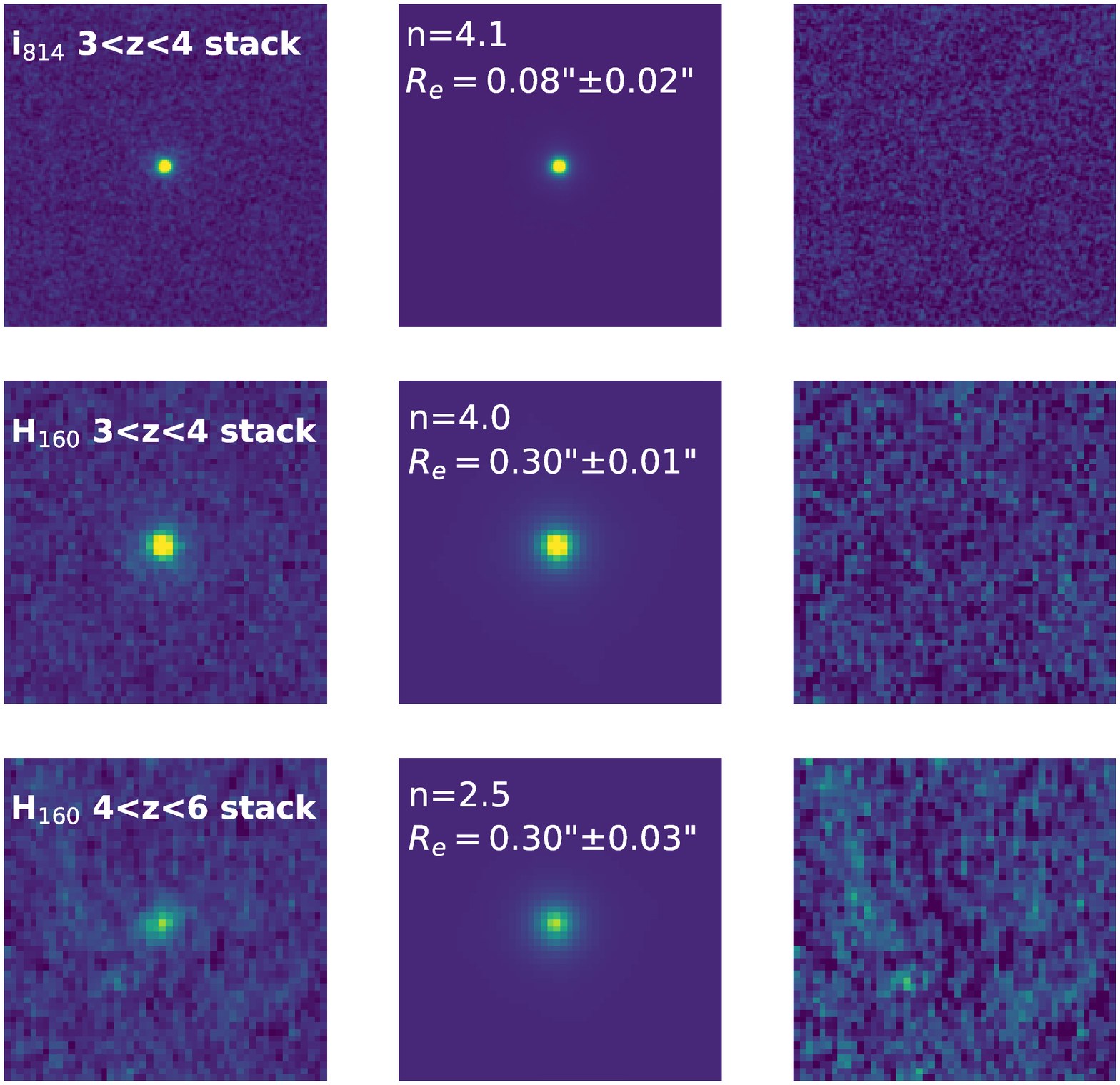}
\caption{GALFIT modelling results. Stamps are $5''\times5''$. 
\textit{Columns} show \textit{Left:} input images, \textit{middle:} single-component S\'ersic model,  \textit{right:} residual of fit model. 
\label{fig-galfits}}
\end{center}
\end{figure}

 
\bibliographystyle{aasjournal}
\bibliography{references}

\end{document}